\documentstyle[prd,aps,floats]{revtex}
\input{psfig.tex}
\newcommand{\beq}{\begin{equation}}
\newcommand{\eeq}{\end{equation}}

\tighten
\begin{document} 
\draft
%\twocolumn[\hsize\textwidth\columnwidth\hsize\csname@twocolumnfalse\endcsname
%DONT FORGET ``]'' below when de-commenting the above line

\preprint{Imperial/TP/97-98/?, DAMTP-1998-139}
\title{Cosmic structure formation in Hybrid Inflation models}
\author{Richard A. Battye{$^{1}$} and Jochen Weller{$^{2}$}}
\address{
${}^1$ Department of Applied Mathematics and Theoretical Physics, University of
Cambridge, \\ Silver Street, Cambridge CB3 9EW, U.K. \\ 
${}^2$ Theoretical Physics Group, Blackett Laboratory, Imperial
College, Prince Consort Road, London SW7 2BZ, U.K.}
\maketitle
\begin{abstract}
%\narrowtext
A wide class of inflationary models, known as Hybrid Inflation models, may
produce topological defects during a phase transition at the end of the
inflationary 
epoch. We point out that, if the energy scale of these defects is close to that
of Grand Unification, then their effect on cosmic structure formation and the
generation of microwave background anisotropies cannot be ignored. Therefore,
it is possible for structure to be seeded by a combination of the adiabatic
perturbations produced during inflation and active isocurvature perturbations
produced by defects. Since the two mechanisms are uncorrelated the power
spectra can be computed by a weighted average of the individual
contributions. We investigate the possible observational consequences of this
with reference to general Hybrid Inflation models and also a specific model
based on Supergravity. These mixed perturbation scenarios have some novel
observational consequences and these are discussed qualitatively.  
\end{abstract}

\date{\today}

%]

\renewcommand{\thefootnote}{\arabic{footnote}}
\setcounter{footnote}{0}

\section{Introduction}
\label{sec-intro}

The precise origin of cosmic structure is one of the most important questions
facing cosmology today. Over the past fifteen years there have been two
competing paradigms: quantum fluctuations created during inflation~\cite{inf}
--- a period of rapid expansion of the universe just after the Planck epoch
which can solve the horizon and flatness problems of the standard Hot Big Bang
model --- and perturbations generated by the gravitational effects of a network
of topological defects~\cite{kib1,VS,HK,zeld,vil}, which may have formed during
some cosmological phase transition close to the energy scale of a Grand
Unification Theory (GUT). In case of inflation the fluctuations are generally
adiabatic, Gaussian and passive in the sense that once created they evolve in a
deterministic way right up to the present day. These assumptions have
simplified the process of making predictions in these models to the point where
accurate $(\sim 1\%)$ calculations of the anisotropies in the cosmic microwave
background (CMB) and the density fluctuations in cold dark matter (CDM) can be
made for a given set of parameters in less than a minute on a modern
workstation~\cite{cmbfast}.  

Making the predictions of the same level of accuracy for defect based models is
much more difficult since the perturbations are isocurvature, non-Gaussian and
are created actively throughout the whole history of the universe, from the
time of defect formation to the present day. However, recent
work~\cite{PSelTa,ABRa,ABRb,ACDKSS,CHMa}, has established a basis for future
work on this subject defining what can be thought of as the standard model,
although there still appears to be some room for understanding more subtle
effects~\cite{CHMa,ABRc}. It was suggested in refs.~\cite{ABRa,ABRb} that flat
universe models with a critical matter density $(\Omega_{\rm m}=1)$ normalized
to COBE would require unacceptably large biases $(\approx 5)$ between cold and
baryonic matter on $100h^{-1}{\rm Mpc}$ scales to be consistent with the
observed galaxy distribution, but more acceptable models can be constructed in
an open universe or one dominated by a cosmological
constant~\cite{ACM,BRA,ASWA}, albeit with a bias of 2 relative to IRAS which
are usually assumed to be good tracers of the underlying mass distribution.
Since COBE normalized adiabatic models based on inflation have no 
problem producing the requisite amount of power on these scales, this suggests
--- if the data is shown to be accurate --- that such models may at least be
partially responsible for the formation of structure. 
 
The idea of combining these two paradigms is a simple one since they are far
from being mutually exclusive; very simply, if the inflationary reheat
temperature is greater than the GUT scale then the post-inflationary universe
will encounter phase transitions, which may form topological defects. More
speculatively, one might form defects in a non-thermal phase transition induced
by parametric resonance~\cite{KLS,TKKL} during the reheating phase after
inflation. But most cosmologists would prefer for there to be only a single
source of fluctuations, based on some kind of `minimalist' principle, and would be sceptical of any theory which has both without further motivation. There
are, however, a wide class of inflationary models, which may produce
topological defects --- usually assumed to be strings, although it is also
possible to produce other kinds of defects --- during a phase transition which
marks the end of the inflationary epoch.  These are known as Hybrid Inflation
models~\cite{hybrid}. Hence, there is sufficient motivation to consider mixed
perturbation scenarios in which structure is formed by both adiabatic density
fluctuations produced during inflation and active isocurvature perturbations
created by defects, without breaking any principle of minimalism, and this is
the subject of this paper. 

In section~\ref{sec-ind} we will discuss the individual components --- the
fluctuations generated by inflation and defects, in particular strings. The
fact that there is no universal model of inflation makes it difficult to make
very specific predictions. Therefore, we will first treat Hybrid Inflation
models in generality by reference to a simple model (see ref.~\cite{LR} for a
compendium of inflationary models --- both Hybrid and otherwise), before
discussing a specific model which was put forward recently to produce inflation
in the context of Supergravity~\cite{LinR}. We will concentrate specifically on
mixing inflation with strings, since they are probably the most obvious
candidate in these scenarios, but most of the general comments that we will
make apply equally well to the case of other topological defects, for example,
the global defect models considered in ref.~\cite{PSelTa}. The models for
strings that we will use are based on those already used in
refs.~\cite{ABRa,ABRb,ABRc,BRA} and we will make two assumptions. Firstly, we
will make the simple assumption that the strings evolve in a perfect scaling
regime, from their formation to the present day, and then we will attempt to
incorporate the effects radiation-matter transition by use of the velocity
dependent one scale model~\cite{MS}. We should note that it is not our
intention in choosing these particular models for inflation and strings to make
any very specific predictions or claims as to their universal validity. Rather,
we wish to discuss qualitatively the sort on phenomena one might possibly
expect in the power spectra and their relation to the current and future
observational data. 

We will then discuss how the spectra can be combined in
section~\ref{sec-obs}. This is in fact trivial since the power spectra should
be uncorrelated and hence the two can be combined by a weighted
average. Clearly, the addition of this extra degree of freedom weakens any
constraint that current observations place on each of the individual models and
we will discuss this in four different contexts. 

Firstly, we consider general Hybrid Inflation models in which the relative
amplitude of the adiabatic and string induced components is arbitrary, along
with the spectral index (the initial density fluctuations created during
inflation are normally assumed to have a simple power law form $P(k)\propto
k^{n}$, where $n$ is the spectral index). Specifically, we will comment on the
constraints which come from analysis of the spectrum of CMB observations on
large scales detected by COBE and also from their combination with measurements
of the density fluctuations on small scales, which are normally quantified in
terms of $\sigma_8$, the fractional over density in spheres with radius
$8h^{-1}{\rm Mpc}$. These are considered to be the most accurate and robust
measurements in cosmology. Comparison to just the COBE data constrains the
spectral index to be in the range $|n-1|<0.2$~\cite{gorski}, while a simple
comparison of the amplitude of the CMB anisotropies with that of $\sigma_8$
rules out the standard CDM scenario with $n=1$ (see, for example,
ref.\cite{BW}), since the COBE normalized value of $\sigma_8$ computed for this
model is approximately twice that which is observed, $\sigma^{\rm OBS}_8\approx
0.6$ in a critical density universe whereas $\sigma^{\rm CDM}_8\approx 1.2$,
favouring a lower value of $n\approx 0.8$. But a more detailed joint
analysis of all the available CMB data and measurements of
$\sigma_8$~\cite{BJK,BJ} suggests that something close $n=1$ gives a better fit
to all the available data. Without performing a full likelihood analysis, we
show qualitatively that these constraints can be relaxed since the large angle
CMB can be induced by strings, allowing for higher spectral indices to fit the
data usually at the expense of the reducing the power on large scales. We will
also consider models which use other cosmological parameters to fit the
measurements of galaxy clustering on large scales ($\approx 50-100h^{-1}{\rm
Mpc}$). We should note that there still remains a strong upper limit on blue
spectra since large spectral indices lead to the production of unacceptable
numbers of primordial black holes~\cite{CGL}. 

We will then discuss the observational aspects of the specific model based on
Supergravity which is introduced in section~\ref{sec-linrio}. In this case the
relative normalization of the adiabatic and string induced components, and the
spectral index, which in this case is also a function of scale, are fixed by a
single parameter of the model. First, we show how the inclusion of the
string component  allows the more extreme values of this parameter, which give
very blue spectra on large scales, to be more compatible with the relative
amplitude of the COBE measurements and those of $\sigma_8$, than if it was
absent. Then we show that simple modifications to the cosmological parameters
can improve the fit to the shape of the observed matter power spectra on large
scales. 

Most speculatively, we examine the possibility that there may be interesting
effects in the power spectrum on small scales. It has been
suggested~\cite{Pa,Pb,SKGPH} that there is a feature in the power spectrum with
wavenumber $k\approx 0.1h{\rm Mpc}^{-1}$, where the Hubble constant is given by
$H_0=100h\,{\rm km}\,{\rm sec}^{-1}\,{\rm Mpc}^{-1}$, and such a feature in the
power spectrum naturally occurs in these models, although not necessarily on
these scales. We illustrate this possibility by reference to a number of simple 
examples, suggesting that the forthcoming redshift surveys (the Sloan Digital
Sky Survey (SDSS) and 2Df) should allow us to test this possibility more
accurately. This feature in the matter power spectrum leads to more power on small scales than in pure adiabatic models and hence it might possible to effect the formation of damped Lyman-$\alpha$ systems and other early objects. We will discuss this in the context of the popular cold plus hot dark matter model which is thought to under produce such features. 

The main focus of our discussion is to reconcile the amplitude
of the COBE detection with the amplitude and shape of the observed galaxy
distribution, a problem which both the standard CDM and defect models both
suffer from. However, the near future will see an explosion in measurements of
the CMB anisotropies over a wide range of scales, for example, from the MAP and
PLANCK satellites. To this end, we finally discuss the implications for the CMB
angular power spectra and the novel features which these models have, in particular the Doppler peak structure and non-Gaussianity. We will focus on the need to exclude or constrain these mixed perturbation scenarios. 

\section{The individual fluctuation spectra}
\label{sec-ind}

\subsection{General hybrid inflation models}
\label{sec-genhybrid}

The proto-typical model for Hybrid Inflation is one which includes two scalar
fields $\phi$, a real scalar field known as the inflaton, and $\psi$, a complex
scalar field which is coupled to the inflaton. The specific potential usually
used is~\cite{hybrid} 
\begin{equation}
V(\phi,\psi)={1\over 4\lambda}\left(M^2-\lambda|\psi|^2\right)^2+{1\over
2}m^2\phi^2+{1\over 2}g^2|\psi|^2\phi^2\,,  
\end{equation}
where $\lambda$ and $g$ are dimensionless coupling constants, and $M$ and $m$
are the mass scales introduced; in particular $M$ is that associated with
spontaneous symmetry breaking, which in the case of $\psi$ being a complex
scalar field leads to the production of global strings at the end of
inflation.

The massive part of the field $\psi$ has an effective mass
$M(\psi)^2=g^2\phi^2-M^2$ and therefore 
for $\phi > \phi_{\rm c}=M/g$ there is a single minimum of the potential in the
$\psi$-direction at $\psi=0$, whereas for $\phi<\phi_{\rm c}$ the potential
develops minima with $|\psi|=M/\sqrt{\lambda}$. In the case where  
\begin{equation}
M^2 \gg  M_{\rm p}m\sqrt{\lambda}\,,\quad M^2\gg m^2/g^2
\end{equation}
and $M_{\rm p}$ is the Planck mass, inflation takes place for $M_{\rm p}>\phi >
\phi_{\rm c}$, with the expansion being dominated 
by the vacuum energy $V(0,0)=M^4/4\lambda$, rather than the false
vacuum. Hence, if the inflaton starts at around $\phi\approx M_{\rm p}$ as in
the Chaotic Inflation scenario~\cite{chaotic}, then inflation takes place as it
rolls down to $\phi_{\rm c}$, at which point the field $\psi$ falls down into
the vacuum manifold creating strings. Of course the universe may continue to
inflate after this point, at least partially diluting the defects, but if the
phase transition takes place sufficiently late, which can always be arranged by
an appropriate choice of the parameters, for example, by ensuring that $M$ is
greater than the Hubble parameter during inflation, then one will be left with
a network of defects which will subsequently evolve toward a scaling regime.  

The adiabatic density perturbations created in this model on a length scale $l$
are given by~\cite{hybrid} 
\begin{equation}
{\delta\rho\over\rho}={2\sqrt{6\pi} g M^5\over 5\lambda\sqrt{\lambda}M_{\rm
p}^3  m^2} 
\left({l\over l_{\rm c}}\right)^{-\beta^2}\,,
\end{equation}
where $l_{\rm c}$ is the horizon size when the inflaton has value
$\phi_{\rm c}$, $\beta=m/\sqrt{3}H$ and  
\begin{equation}
H^2={2\pi M^4\over 3\lambda M_{\rm p}^2}\,,
\end{equation} 
is the Hubble parameter when $\phi=\phi_{\rm c}$. This model has a spectral
index $n=1+2\beta^2>1$, and therefore if $H\gg m$ then the spectrum is almost
scale free, whereas if $m\sim H$ then the spectral index can be much larger than
one. In ref.\cite{hybrid} two possible scenarios were considered. Firstly, if
$g^2\sim\lambda\sim 10^{-1}$ and $m\sim 10^2{\rm GeV}$, then normalization to
the observed fluctuations [($\delta\rho/\rho)_{\rm obs}\sim 5\times 10^{-5}$,
in these units] yields $M\sim 10^{11}{\rm GeV}$, and hence the adiabatic
density fluctuations will dominate over those produced by the strings
[($\delta\rho/\rho)_{\rm str}\sim GM^2\sim 10^{-10}(\delta\rho/\rho)_{\rm
obs}$]. In this case the creation of strings at this scale may have other
interesting cosmological implications --- the production of dark matter axions
by the radiative decay of axions strings~\cite{BatShe} --- but they will not
have a substantial effect on structure formation. However, if $g\sim\lambda\sim
1$ and $m\sim 5\times 10^{10}{\rm GeV}$ then normalization requires that $M\sim
10^{15}{\rm GeV}$ --- around the GUT scale --- with a spectral index of
$n\approx 1.1$. In this case the perturbations created by the strings must be
taken into account since they are of comparable size to the adiabatic ones. 

Since this simple model was first proposed, there have been many other Hybrid
Inflation models discussed in the literature with a wide variety of predictions
for the energy scale and spectral index, most of which adhere to the dogma that
the natural scale for inflation is below the GUT scale and the perturbations
are scale free (although see ref.~\cite{GBL}). However, this simple example
stands as a illustration that simple models which create topological defects at
the end of inflation exist and moreover the defects can be sufficiently massive
to seed density perturbations which are of comparable size to the adiabatic
perturbations created during inflation. In the next section we will discuss a
particular model for Hybrid inflation based on Supersymmetry and Supergravity
which makes specific parameter based predictions for the energy scale and
spectrum of initial fluctuations. These predictions will be used to illustrate
the novel observational features of these mixed perturbation scenarios. We
should, however, be mindful of the fact that no universally accepted model of
inflation exists and many more models will be invented to try to reconcile the
theoretical and observational prejudices of the day. Although many models will
not have GUT scale defects as a generic consequence, it is a very natural scale
for symmetry breaking transitions to occur and models which predict this are
well worth investigating, even if their observational consequences appear, at
first glance, to be at variance with the observations. In
section~\ref{sec-genhybridstrings} we will illustrate the effects of simple
models with an arbitrary normalization, close to the GUT scale, and spectral
indices which are not one, showing that, at least qualitatively, that they may
be to account for some of the current observations.   

\subsection{A specific inflation model based on Supergravity}
\label{sec-linrio}

In the previous section we discussed the original Hybrid Inflation model,
illustrating its salient features from the cosmological point of view. However,
such a model does not have any particular motivation from the point of view of
fundamental physics --- the ultimate aim of these endeavours. Hence, for
definiteness, we would like to have a specific model which has much stronger
origins in the realm of high energy physics theories such as
Supergravity. There are a number of popular models which use the flat
directions of superpotentials to allow slow-roll inflation, notably
the F-term~\cite{fterm} and D-term~\cite{dterm} scenarios, some of which also
produce topological defects during the phase transition at the end of inflation
and could be candidates for the ideas that we are discussing here. However, we have chosen a model introduced by Linde and Riotto~\cite{LinR} to illustrate the
qualitative nature of these scenarios. Although there are some subtle
philosophical problems associated with this model~\cite{LR}, for specific
choice of the parameters it exhibits a number of the properties which we are
interested in investigating. Here, we will derive the initial fluctuation
spectrum for this model, following and extending the original work~\cite{LinR},
and then we will incorporate the initial spectra into CMBFAST, the standard
linear Einstein-Boltzmann solver, to compute the spectrum of CMB anisotropies
and CDM. From now onwards we will work in units where the Planck mass
appropriate for high energy physics applications, such as String Theory, is
given by $(8\pi G)^{1/2}=1$. 

The model is based on the simplest choice of the superpotential which takes the
form~\cite{CLLSW} 
\beq
W=S\left(\kappa\bar{\phi}\phi-\gamma^2\right)\,,
\eeq
where $\phi$ and $\bar{\phi}$ are conjugate pairs of chiral superfields, $S$ a
neutral superfield and $\kappa<1$ is a parameter of the theory. If we
canonically normalize the scalar field $S=\sigma/\sqrt{2}$, then the effective
potential, excluding the D-term, is given by 
\beq
V={\kappa^2|\sigma|^2\over 2}\left (
|\phi|^2+|\bar\phi|^2\right)+|\kappa\bar\phi\phi-\gamma^2|^2\,. 
\eeq
For $\sigma>\sigma_{\rm c}=\gamma\sqrt{2/\kappa}$, the mass of the chiral
superfields is positive and the potential in their directions has a single
minimum at $\phi=\bar\phi=0$. Along this direction the potential is flat and
so a viable inflationary model cannot be constructed. However, if one makes a
slight modification to the potential it is possible, for example, by softly
breaking Supersymmetry~\cite{CLLSW}. The novel aspect of the Linde-Riotto model
is that they suggested this may be done by the inclusion of radiative and
Supergravity corrections. In this case, if the field $\sigma$ starts at some
large value as in the Chaotic Inflation scenario and the original Hybrid
Inflation model, then inflation takes place until the field rolls down  to
$\sigma=\sigma_{\rm c}$, at which point the field will quickly roll down to the
absolute minimum of the potential where $\sigma=0$ and
$\phi=\bar\phi=\gamma/\sqrt{\kappa}$, producing topological defects if the
vacuum manifold has non-trivial homotopy. It was suggested in ref.~\cite{LinR}
that such models can naturally have the correct symmetry breaking schemes to form cosmic strings and that the energy scale of these strings is given by
$\gamma/\sqrt{\kappa}$ which for an appropriate choice of parameters could be
the GUT scale. 

In order to compute the potential, therefore, we must include radiative effects
and Supergravity corrections. The one loop SUSY potential for this model is
given by~\cite{DSS,LSS}  
\beq 
V_1 =
\displaystyle\frac{\kappa^2}{128\pi^2}\left[\left(\kappa\sigma^2-2\gamma^2\right)^2\ln\frac{\kappa\sigma^2-2\gamma^2}{\Lambda^2}+\displaystyle
\left(\kappa\sigma^2+2\gamma^2\right)^2\ln\frac{\kappa\sigma^2+2\gamma^2}{\Lambda^2}-2\kappa^2\sigma^4\ln\frac{\kappa\sigma^2}{\Lambda^2}\right]\,,
\eeq
where $\Lambda$ is the renormalization scale which is introduced in the usual
way. Since the period of inflation that we are interested in takes place for
$\sigma\gg\sigma_{\rm c}$, one can expand this correction to the potential as 
\beq
V_1=\gamma^4\frac {\kappa^2}{8\pi^2}\ln\frac{\sigma}{\sigma_c}+\cdots\,.
\eeq
If the field is very large at the beginning of inflation then Supergravity
corrections will also be important. The SUGRA correction to the potential is
given by 
\beq
V_{\rm
SUGRA}=\displaystyle\gamma^4\exp\left(\frac{\sigma^2}{2}\right)\left[1-\frac{\sigma^2}{2}+\frac{\sigma^4}{4}\right]-\gamma^4=\displaystyle\gamma^4
\frac{\sigma^4}{8}+\cdots\,, 
\eeq
and hence by combining the two one gets the fully corrected approximate
potential    
\begin{equation}
V=\gamma^4\left(1+\frac{\kappa^2}{8\pi^2}\ln\frac{\sigma}{\sigma_c}+\frac{\sigma^4}{8}\right)\,,
\end{equation}
which is strictly valid for $\sigma\gg\sigma_{\rm c}$, but should also be
useful for smaller values of $\sigma$ close to $\sigma_{\rm c}$. 

Using this potential one can compute the evolution of the scalar field, which
we have done in terms of $N$ the number of $e$-foldings until the end of
inflation. During the inflationary epoch $3H^2=V$ and
$3H\dot\sigma=-V^{\prime}$, from which we can deduce that
$\dot\sigma/H=-V^{\prime}/V$ and hence  
\beq 
{d\sigma\over dN}={V^{\prime}\over V}\,,
\eeq
since the scale factor $a\propto e^{-N}$. If we assume that the potential does
not change appreciably during the early stages of inflation when the
cosmologically interesting perturbations are being created, then one can solve
the equation for the field $\sigma$ in terms of $N$, 
\beq
\sigma^2={\kappa\over 2\pi}{\displaystyle{2\pi\over\kappa}\sigma^2_{\rm
end}+\tan\left({\kappa\over 2\pi}N\right) \over 1
+\displaystyle{2\pi\over\kappa}\sigma^2_{\rm end}\tan\left({\kappa\over
2\pi}N\right)}\,, 
\label{sig}
\eeq
where $\sigma_{\rm end}$ is the value of the field when inflation ends, that
is, when $N=0$. Assuming this value of the field to be relatively small, which
may not always be the case, one can further approximate (\ref{sig}) to give  
\begin{equation}
\sigma
\approx\sqrt{\frac{\kappa}{2\pi}\tan\left(\frac{\kappa}{2\pi}N\right)}\,. 
\label{approxfield}
\end{equation}
From this we can see that there is a constraint on the value of the parameter
$\kappa$ from the requirement that there have been at least 60 $e$-foldings, so
as to make the universe almost flat by the present day. The argument of the
tangent function should never be allowed to be greater than $\pi/2$, and hence
one can deduce that $\kappa<\kappa_{\rm c}=\pi^2/60\approx 0.16$. The 
number of $e$-foldings $N$ can then be related to the wavelength or wavenumber
of the perturbations ($l=2\pi/k$) by 
\beq
N\approx 54 -\ln\left(\frac{k}{1h{\rm Mpc}^{-1}}\right)\approx
52+\ln\left(\frac{l}{1h^{-1}\rm Mpc}\right)\,, 
\eeq
where we have assumed that the size of the universe is $3000h^{-1}{\rm Mpc}$ at the present day and hence we can compute parametrically the field $\sigma$ as a function of the
comoving scale during inflation. 

The spectrum of primordial density fluctuations generated by the evolution of
this field during inflation can be approximated by  
\begin{equation}
\frac{\delta\rho}{\rho}\approx \frac{\sqrt{3}}{5\pi}\frac{V^{3/2}}{V^{\prime}}
\approx 
\frac{\sqrt{3}}{5\pi} \gamma^2
\left(\frac{\kappa^2}{8\pi^2}\frac{1}{\sigma}+\frac{1}{2}\sigma^3\right)^{-1}\,,
\end{equation}
and spectral index, which is now scale dependent, is given by 
\begin{equation}
n-1 \approx 2\frac{V^{\prime\prime}}{V}
=\left(3\sigma^2-\frac{\kappa^2}{4\pi^2}\frac{1}{\sigma^2}\right)\,. 
\label{spectral}
\end{equation}
Clearly, the spectrum is not totally scale invariant since the field evolves
according to (\ref{sig}) during inflation and in general the effective spectral
index increases as one approaches the largest scales, but we now have a simple
parametric formula which allows us to compute the initial spectrum of density
fluctuations $P_{\rm i}(k)$. One might also expect such models to create a
tensor contribution to the CMB fluctuations on large scales, but this would be
proportional to $V^{1/2}$ and hence $(\delta T/T)_{\rm T}={\cal O}(1)\gamma^2$,
which in the regime under consideration here is much smaller than that due to
scalars~\cite{linde}. 

Assuming that these adiabatic perturbations are the only contribution to the
scales probed by COBE, one can use the normalization~\cite{lyth}  
\beq
5.3\times 10^{-4}\approx\displaystyle{V^{3/2}\over
V}\bigg{|}_{\sigma=\sigma_{60}}\approx\gamma^2\sigma_{60}\left(\displaystyle{\kappa^2\over
8\pi^2}+{1\over 2}\sigma^4_{60}\right)^{-1} 
\eeq
where $\sigma_{60}$ is the value of the inflaton field when the current
observable universe leaves the horizon during inflation. Therefore, using
(\ref{approxfield}) one can deduce that 
\beq
1.7\times
10^{-5}\approx\displaystyle{\gamma^2\over\kappa^{3/2}}\left[\sin\left({30\kappa\over\pi}\right)\right]^{1/2}\left[\cos\left({30\kappa\over\pi}\right)\right]^{3/2}\,,
\eeq
and hence if $\kappa$ is small, for example $\kappa=0.08$, then the symmetry breaking scale for the strings is given by $\gamma/\sqrt{\kappa}\approx 2.3\times 10^{-3}$. If the phase transition
which takes place at the end of inflation leads to the formation of strings,
then their mass per unit length will be  
\beq
G\mu\approx {\gamma^2\over 8\pi\kappa}\approx 2.1\times 10^{-7}\,.
\eeq
Whereas, if $\kappa$ is close to $\kappa_{\rm c}$, for example $\kappa=0.15$,
then one can deduce that $G\mu\sim 5 \times 10^{-6}$. In both cases the
strings are around the scale at which they will also contribute substantially
to the the COBE normalization and clearly they must be taken into account. In
section~\ref{sec-super}, we will discuss the implications of this for the
relative contribution of adiabatic and string induced perturbations, and hence
the computed spectra. For the moment we will ignore the strings and compute the
spectrum of CMB anisotropies and fluctuations in the CDM, assuming that the
adiabatic fluctuations are the only contribution. 

\begin{figure}
\setlength{\unitlength}{1cm}
\centerline{\psfig{file=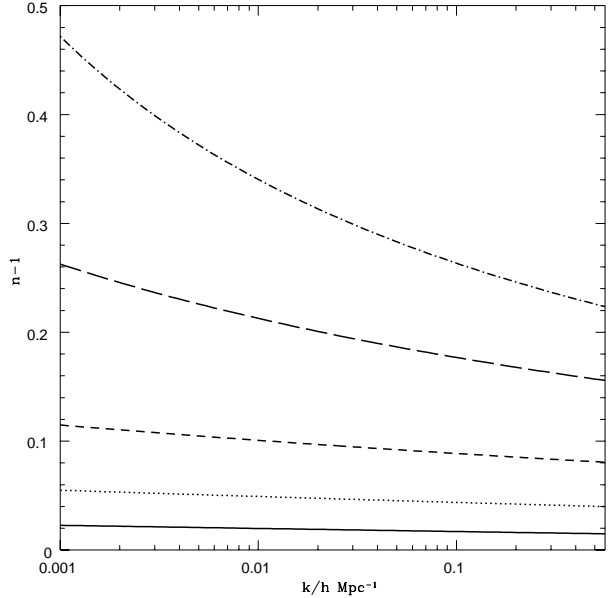,width=3.4in}}
\caption{The effective spectral index as function of the comoving scale during
inflation for $\kappa=0.08$ (solid line), $\kappa=0.1$ (dotted line),
$\kappa=0.12$ (short dash line), $\kappa=0.13$ (long dash line) $\kappa=0.14$
(dot-short dash line) and $\kappa=0.15$ (dot-long dash line). } 
\label{fig-spectral}
\end{figure}

The spectrum of initial fluctuations in this model is not a simple power law,
and hence the spectral index is now a function of $k$, that is, $P_{\rm
i}(k)\propto k^{n(k)}$. The effective spectral index is plotted in the
cosmologically interesting range of $k$  
for various values of $\kappa$ in fig.~\ref{fig-spectral}. For $\kappa\le
0.13$, the spectral index is approximately constant, but for larger values of
$\kappa$ the spectrum rises sharply on large scales (small $k$). This novel
feature of the spectrum makes this model particularly interesting from the
point of view of this paper since it will lead to a heavily blue shifted
spectrum on very large scales and such a spectrum is tightly constrained by the
observations. However, if the large angle part of the CMB spectrum is created
by the strings then it may yet be possible for this model to be viable.  

\begin{table}[!t]
\begin{center}
\begin{minipage}{8.0cm}
\begin{tabular}{ccccccc}
%\hline 
$\kappa$ & 0.08 & 0.10 & 0.12 & 0.13 & 0.14 & 0.15 \\
\hline
$\sigma_8$ & 1.26 & 1.37 & 1.61 & 1.85 & 2.39 & 4.04 \\
%\hline
\end{tabular}
\end{minipage}
\end{center}
\medskip
\caption{The computed values of $\sigma_8$ for the Linde-Riotto model of
inflation using the standard cosmological parameters, and a varying the
parameter $\kappa$. Note that we have not included the possible effect of
strings at this stage.} 
\label{tab-kappa}
\end{table}

\begin{figure}
\setlength{\unitlength}{1cm}
\begin{minipage}{8.0cm}
\leftline{\psfig{file=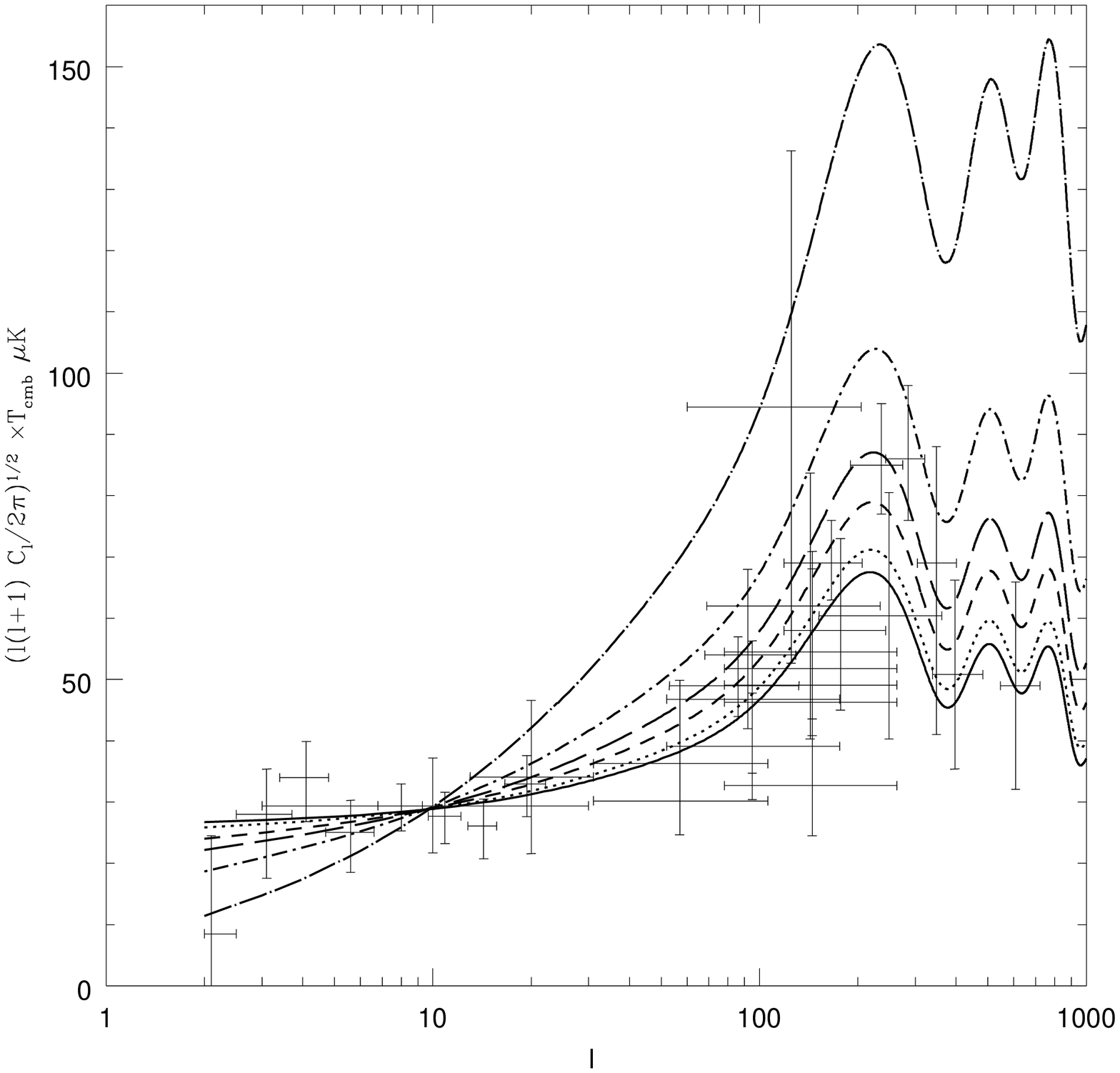,width=3.4in}}
\end{minipage}\hfill
\begin{minipage}{8.0cm}
\rightline{\psfig{file=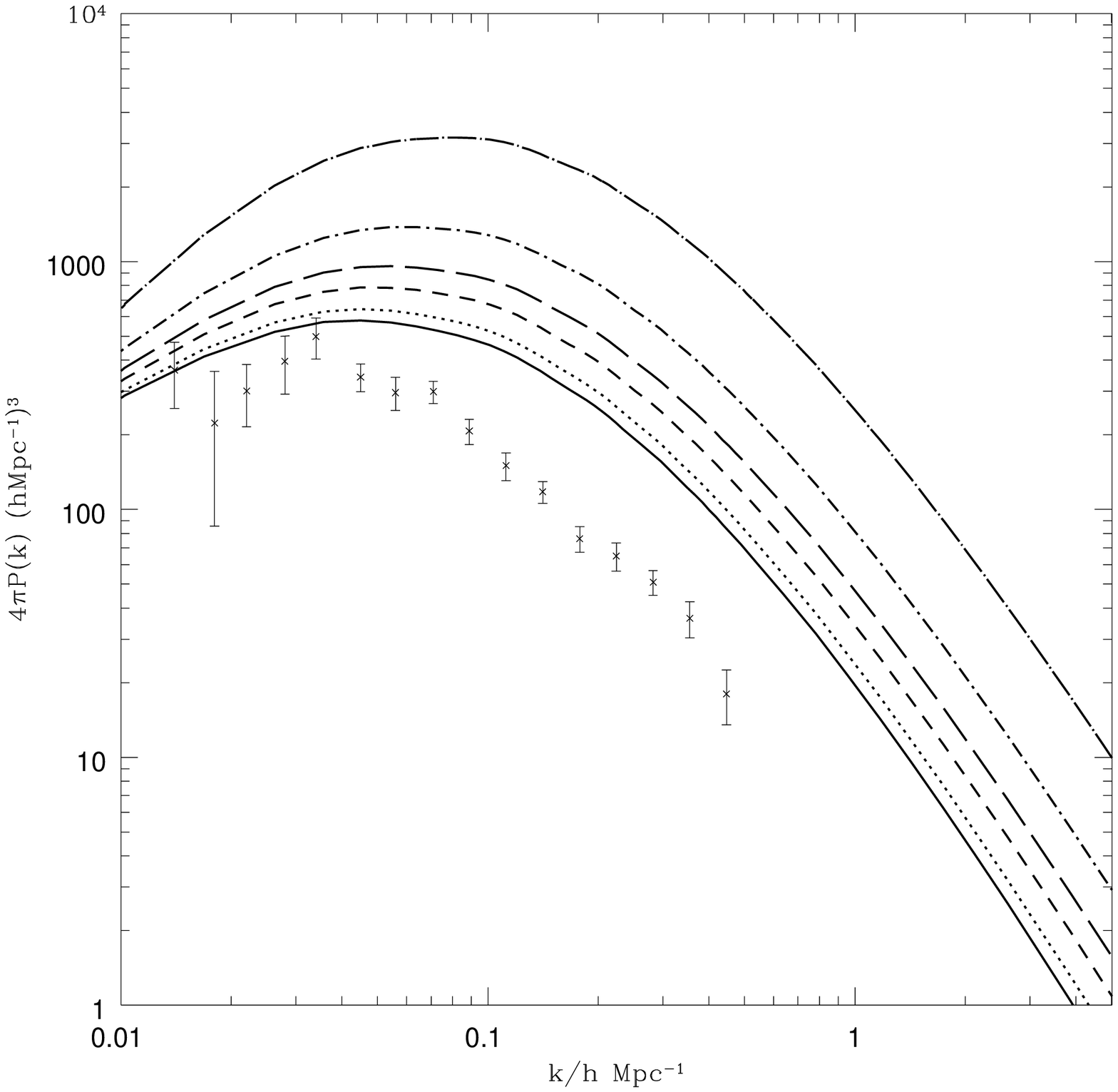,width=3.4in}}
\end{minipage}
\caption{On the left the angular power spectrum of CMB anisotropies and on the
right the power spectrum of fluctuations in the CDM for the Linde-Riotto model without a string induced component, 
using the standard cosmological parameters and $\kappa=0.08$ (solid line),
$\kappa=0.1$ (dotted line), $\kappa=0.12$ (short dash line), $\kappa=0.13$
(long dash line), $\kappa=0.14$ (dot-short dash line) and $\kappa=0.15$
(dot-long dash line). In both cases the current observational data points are
also included to guide the eye. Notice that the CMB anisotropies for
$\kappa=0.15$ are wildly at odds with the observations at all scales and that even
the models with smaller values of $\kappa$ are clearly at odds with the
amplitude and shape of the observed matter power spectrum.} 
\label{fig-varykappa}
\end{figure}

Now that we have derived the power spectrum of initial density fluctuations, it
can be incorporated into CMBFAST in order to compute the power spectra of CMB
anisotropies and matter fluctuations in the CDM which we would observe today
for a given value of $\kappa$. The spectra are presented in
fig.~\ref{fig-varykappa} for various values of $\kappa$ and the standard
cosmological parameters\footnote{The standard cosmological parameters used here
and throughout this paper, unless stated otherwise, are a universe comprizing
of $95\%$ CDM and $5\%$ baryons, with three massless neutrinos, a Hubble
constant $H_{\rm 0}=50\,{\rm km}\,{\rm sec}^{-1}\,{\rm Mpc}^{-1}$ and the
standard recombination  history.}, and the corresponding values of $\sigma_8$
are tabulated in table 1. If $\kappa$ is small (for example, $\kappa\approx
0.08$) then the initial fluctuations are almost scale invariant and the results
are essentially just those for standard CDM, but as $\kappa$ increases the
spectrum develops a tilt toward smaller scales, this being most graphically
illustrated by the extreme case of $\kappa=0.15$ where there are almost exactly
60 $e$-foldings of inflation. When compared to the current
observations~\cite{tegmark,PDa,PDb}, it is clear, just by inspection, that as
they stand the models with larger values of $\kappa$ would be ruled out by the
observations of the CMB  and  galaxy correlations on small-scales. 

\subsection{String models}
\label{sec-string}

We have chosen strings as an example of topological defects produced at the end
of inflation due to the well established property~\cite{kib2,AT,BB,AS} that
they evolve toward a self-similar scaling regime, in which the large scale
properties of the network are described by a single scale and the density
remains constant relative to the horizon. It is this property which makes them
a possible source of an almost scale invariant spectrum of density
perturbations across a wide range of scales, and hence a realistic model for
structure formation. We could, of course, have used other topological defect
models, for example, the global defect models used in ref.~\cite{PSelTa} and
the qualitative predictions would be very similar. We should note that in doing
this we have made the assumption that the initial distribution of strings is
such that the network can achieve  a scaling regime before the cosmologically
interesting scales come inside the horizon. We have already pointed out that if
a substantial period of inflation were to take place after the phase
transition, then the strings would become diluted possibly radically altering
the standard picture of string evolution. But, so long as inflation ends
quickly enough, which is usually possible by tuning parameters, this dilution
can be reversed before the cosmologically interesting epoch, just before the
time of radiation-matter equality.  

\begin{figure}
\setlength{\unitlength}{1cm}
\begin{minipage}{8.0cm}
\leftline{\psfig{file=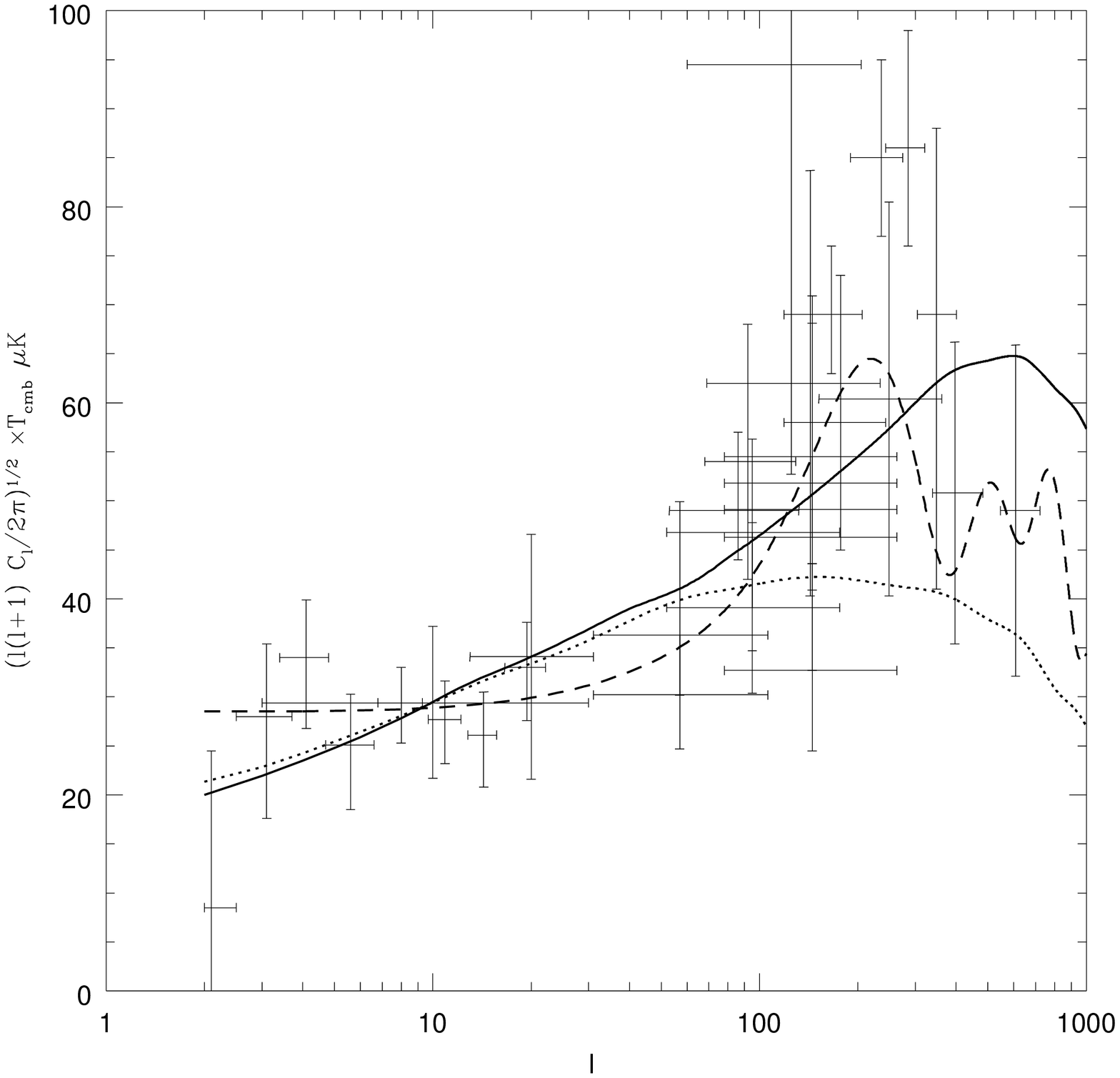,width=3.4in}}
\end{minipage}\hfill
\begin{minipage}{8.0cm}
\rightline{\psfig{file=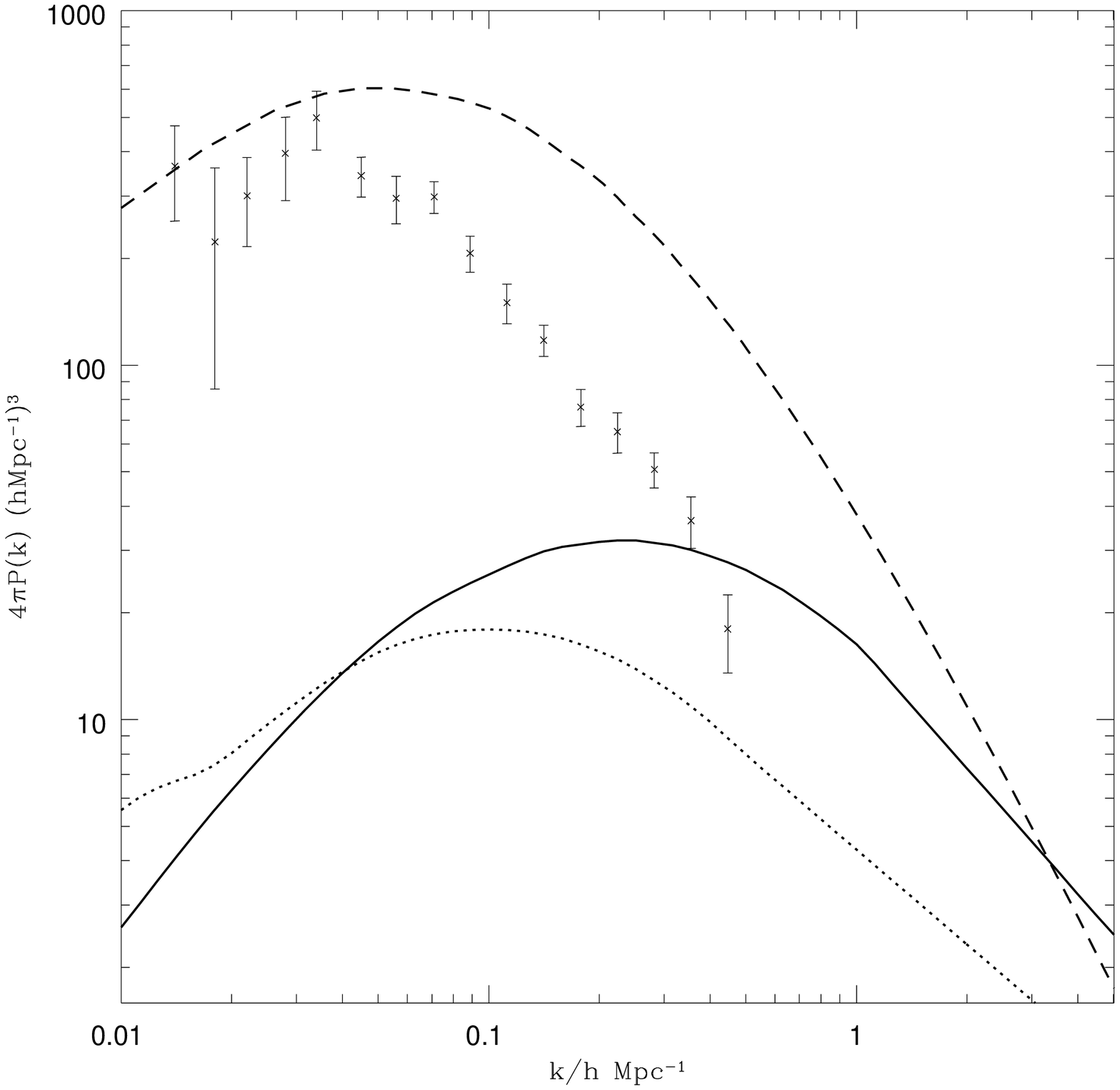,width=3.4in}}
\end{minipage}
\caption{On the left the angular power spectrum of CMB anisotropies and on the
right the power spectrum of fluctuations in the CDM for the standard scaling
source (solid line) and the standard string model (dotted line). In both cases
the standard cosmological parameters have been used. The
current observational data points and the equivalent spectra for the standard
CDM scenario (short dash line) are also included to guide the eye.} 
\label{fig-stdstr}
\end{figure}

The particular model we will use to describe the two-point correlation
functions of the strings is that which was developed in
refs.\cite{VHS,ABRa,ABRb}, where the string network is modelled as an ensemble
of straight segments each with size $\xi\eta$, where $\eta$ is the conformal
time, and a random velocity chosen from a Gaussian distribution which has zero
mean and variance $v$. The scaling regime is usually achieved by the production
of loops and subsequent emission of radiation into the preferred channel,
usually assumed to be gravitational radiation. As a first approximation this
can be accounted for by removing string segments at a rate which exactly
maintains scaling. The results of using this approximation, which we shall call
the standard scaling source, are presented in fig.~\ref{fig-stdstr} along with
some data points which represent the
observations. The main qualitative features of this
model are the apparent absence of any kind of Doppler peak in the CMB
anisotropies and a matter power spectrum which on large scales $(\approx
100h^{-1}{\rm Mpc}$) appears to require a bias of around 5 and the computed
value of $\sigma_8\approx 0.31$. 

In order to model a network of strings more realistically one can do two
things. Firstly, one must attempt to take into account the effects of the
matter radiation transition; scaling is a balance between the rate of expansion
of the universe and the efficiency by which the network can lose energy
into loops. During the transition era, that is, $0.1\eta_{\rm
eq}<\eta<100\eta_{\rm eq}$, where $\eta_{\rm eq}$ is the conformal time of
equal matter-radiation, the expansion rate is relaxing from the radiation era,
where the scale factor is proportional to $\eta$, to the matter era, where it
is proportional to $\eta^2$. Clearly, the nature of the scaling changes during
this time and it has been suggested~\cite{MS} that the change in the density of
strings observed in the two different eras can be modelled using the velocity
dependent one-scale model. This model treats the two parameters, $\xi$ and $v$,
used in construction of the two point functions as being dependent on the
conformal time, allowing one to compute the rate at which the density changes. 

Another aspect of string evolution which is not described by this simple model
is the effect of small-scale structure. In high resolution
simulations~\cite{AT,BB,AS}, it was found that small-scale structure built up
close to the resolution of the simulation due to the copious production of loops on these scales. Although in reality this will be stabilized by radiation
backreaction, some structure will remain effectively renormalizing the mass per
unit length of these string segments to be $\tilde\mu\approx 2\mu$, where $\mu$
is the `bare' mass per unit length. Formally, this can be done by using a
transonic equation of state for the string~\cite{carter}, that is, treating the
strings as having a more complicated equation of state, rather than the usual
Nambu-Goto one, where the energy per unit length and the tension are equal. In
Minkowski space, the energy momentum tensor for a general string is  
\beq
T_{\mu\nu}(x)=\int d\sigma\delta^{4}(x-X(\sigma,t))\left[U{\dot X}_{\mu}{\dot
X}_{\nu}-TX_{\mu}^{\prime}X_{\nu}^{\prime}\right]\,, 
\eeq 
where $X^{\mu}(\sigma,t)$ are the spacetime coordinates of the string at time
$t$ parameterized by $\sigma$, some arbitrary coordinate along the string, $U$
is the energy density of the string and $T$ is its tension. For the special
case of a Nambu-Goto string the equation of state is $T=U=\mu$, but for the
transonic case under discussion here $TU=\mu^2$ and, therefore, if
$U=\tilde\mu\approx 2\mu$ then $T/U\approx 1/4$. By making this simple
modification to the original model, one can incorporate some of the effects of
small-scale structure. In particular, these modifications can lead to an
enhanced peak structure due to the effects of an enhanced Newtonian
potential~\cite{vilpriv}. A detailed investigation of these effects is the
subject of work in progress. 

Using the modifications described above, we have computed the spectra of CMB
anisotropies and fluctuations in the CDM for the standard set of cosmological
parameters  and the results are also presented in fig.~\ref{fig-stdstr}. The
main qualitative features of this model, which we will call the standard string
model, are a slightly titled spectrum of CMB anisotropies which rises to a
single broad peak around $l=400-600$ with no secondary oscillations, and a
matter power spectrum which appears to match the observation extremely badly
both in amplitude and shape\footnote{Both of these deficiencies can be
rectified by the inclusion of a cosmological constant with
$\Omega_{\Lambda}\approx 0.7-0.8$ (see, for example, ref.~\cite{BRA})}. The
computed value for $\sigma_8\approx 0.42$ --- reasonably close to the measured
value --- and that for $G\tilde\mu\approx 2.0\times 10^{-6}$, which implies
that $G\mu\approx 1.0\times 10^{-6}$ once small-scale structure is taken into
account. This well within the constraint imposed by absence of timing residuals
in the observations of milli-second pulsars~\cite{CBS}. 

We should emphasize that the predictions of these two simple assumptions are
not definitive. They appear to have very similar predictions on large
scales, but their predictions on smaller scales at very different, for
example, very different values of $\sigma_8$. At this stage it seems sensible
to consider the implications of both models and hopefully future work will
enable us to pin down the predictions of these scenarios more fully.  

\section{Combining the spectra and its observational consequences}
\label{sec-obs}

\subsection{Combination by a weighted average}
\label{sec-ave}

At first sight it may appear that combining the effects of adiabatic
fluctuations created by inflation and those created actively by topological
defects is a highly non-trivial task and indeed if, for example, one were
trying to create CMB sky maps by considering the evolution of each mode, it
would be. However, computations are simplified considerably by the fact that,
at this stage, we are only trying to compute the power spectra. Since each of
the two sources are uncorrelated --- one happening during the inflationary
epoch and the other after --- one can simple add the correctly normalized
spectra together. 

The only subtle aspect is to normalize each of the two contributions so that
the sum is normalized relative to COBE. If one assumes, for the moment, that
the normalization of each of the components is arbitrary and to be computed
from the observations, one can add the spectra as 
\beq
C^{\rm tot}_{\ell}=\alpha C^{\rm adia}_{\ell}+(1-\alpha)C^{\rm
str}_{\ell}\,,\quad P^{\rm tot}(k)=\alpha P^{\rm adia(k)}+(1-\alpha)P^{\rm
str}(k)\,, 
\eeq
where $0\le\alpha\le 1$ is an arbitrary constant defining the relative
normalization, $C^{\rm adia}_\ell$ and $P^{\rm adia}(k)$ are the spectra from
adiabatic perturbations individually normalized to COBE, and  $C^{\rm str}_\ell$
and $P^{\rm str}(k)$ are those for strings. In a specific high energy physics
motivated model, for example, the one discussed in section~\ref{sec-linrio},
the relative normalization of the two components will be fixed by the
parameters of the model, effectively fixing the value of $\alpha$.  

\begin{figure}
\setlength{\unitlength}{1cm}
\begin{minipage}{8.0cm}
\leftline{\psfig{file=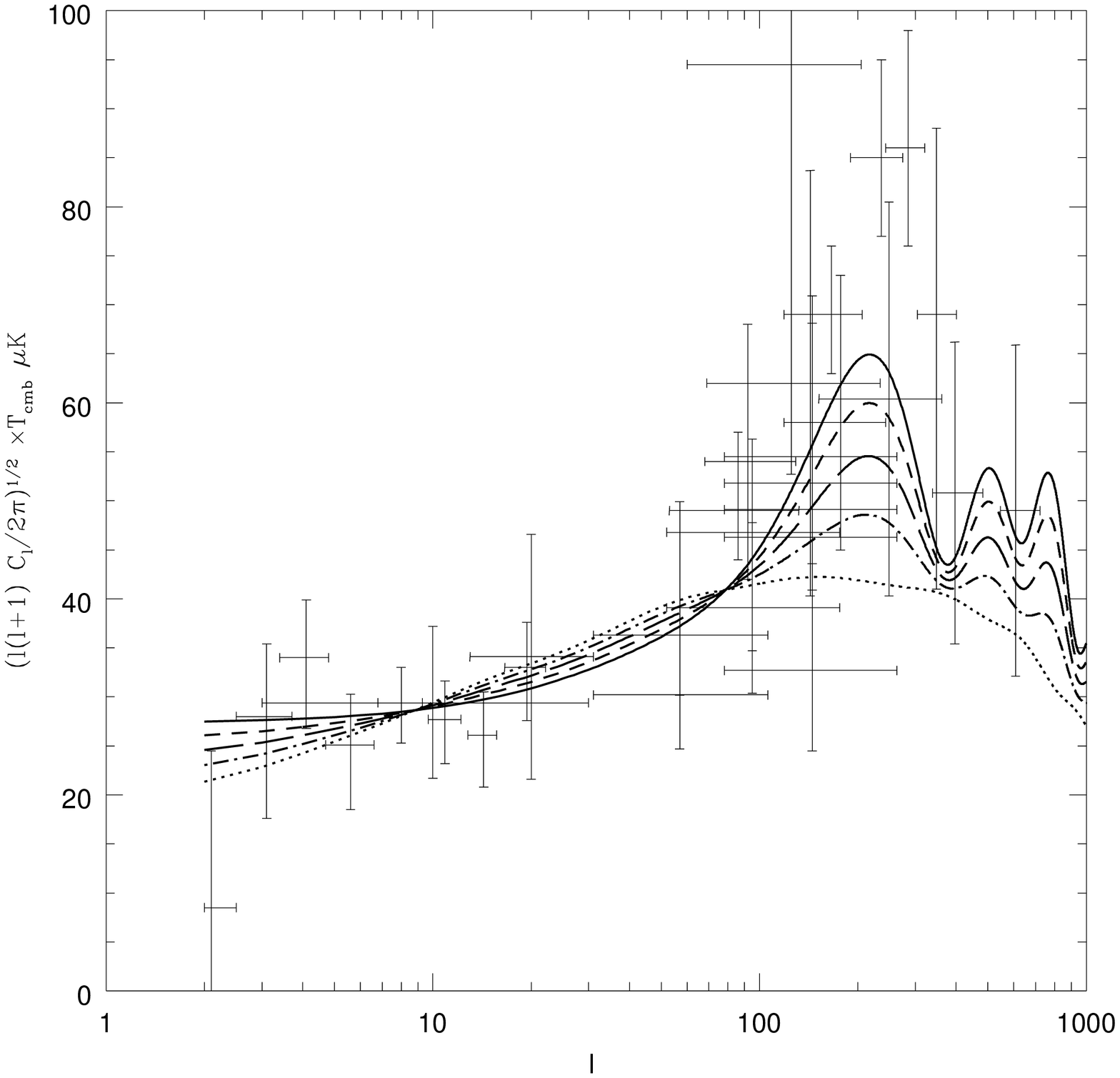,width=3.4in}}
\end{minipage}\hfill
\begin{minipage}{8.0cm}
\rightline{\psfig{file=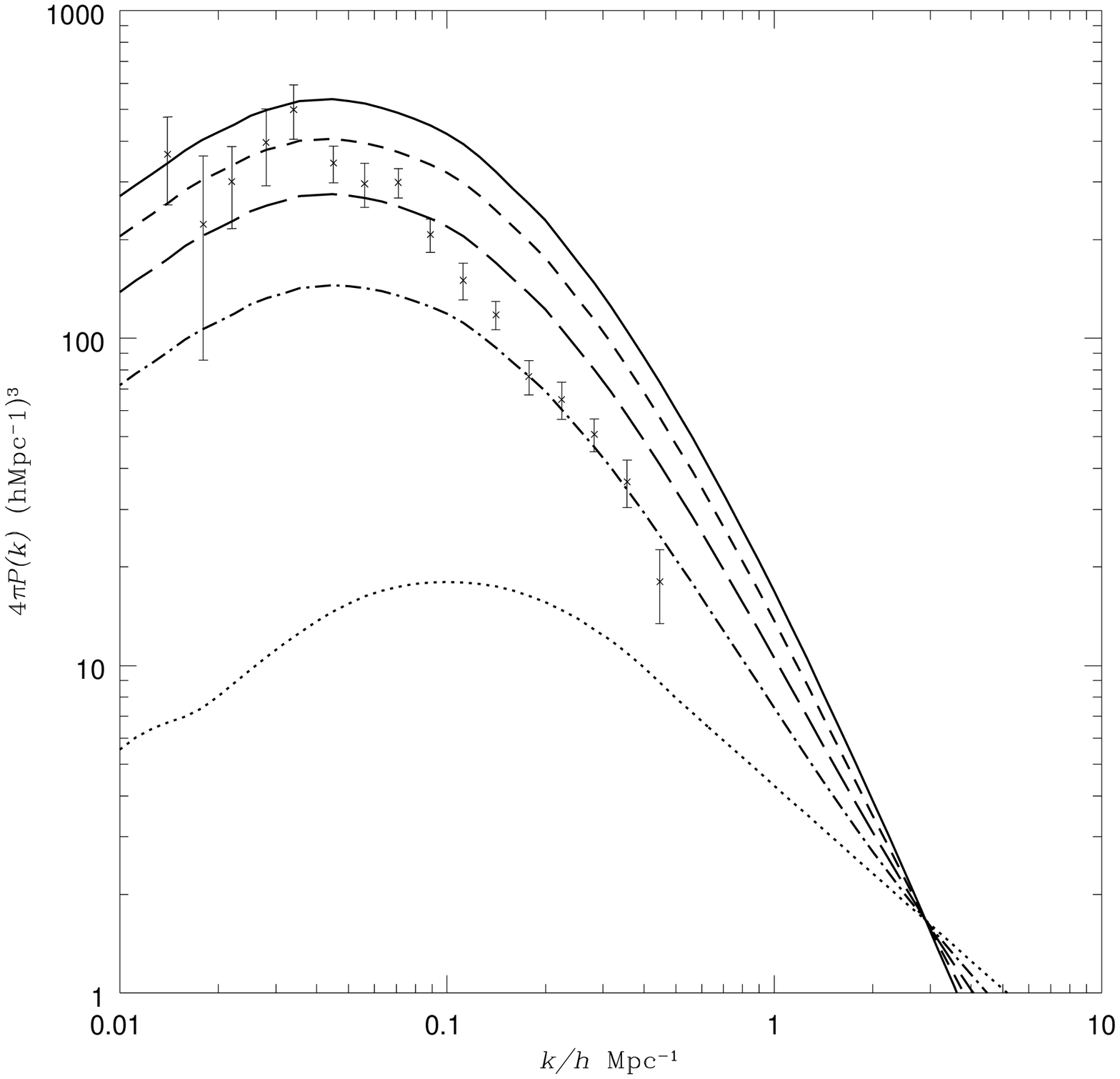,width=3.4in}}
\end{minipage}
\caption{On the left the angular power spectrum of CMB anisotropies and on the
right the power spectrum of fluctuations in the CDM for the standard CDM
scenario mixed with standard scaling source using a ratio of $\alpha=1.0$
(solid line), $\alpha=0.75$ (short dash line), $\alpha=0.5$ (long dash line),
$\alpha=0.25$ (short dash-dotted line) and $\alpha=0.0$ (dotted line).} 
\label{fig-scdmscaling}
\end{figure}

\begin{figure}
\setlength{\unitlength}{1cm}
\begin{minipage}{8.0cm}
\leftline{\psfig{file=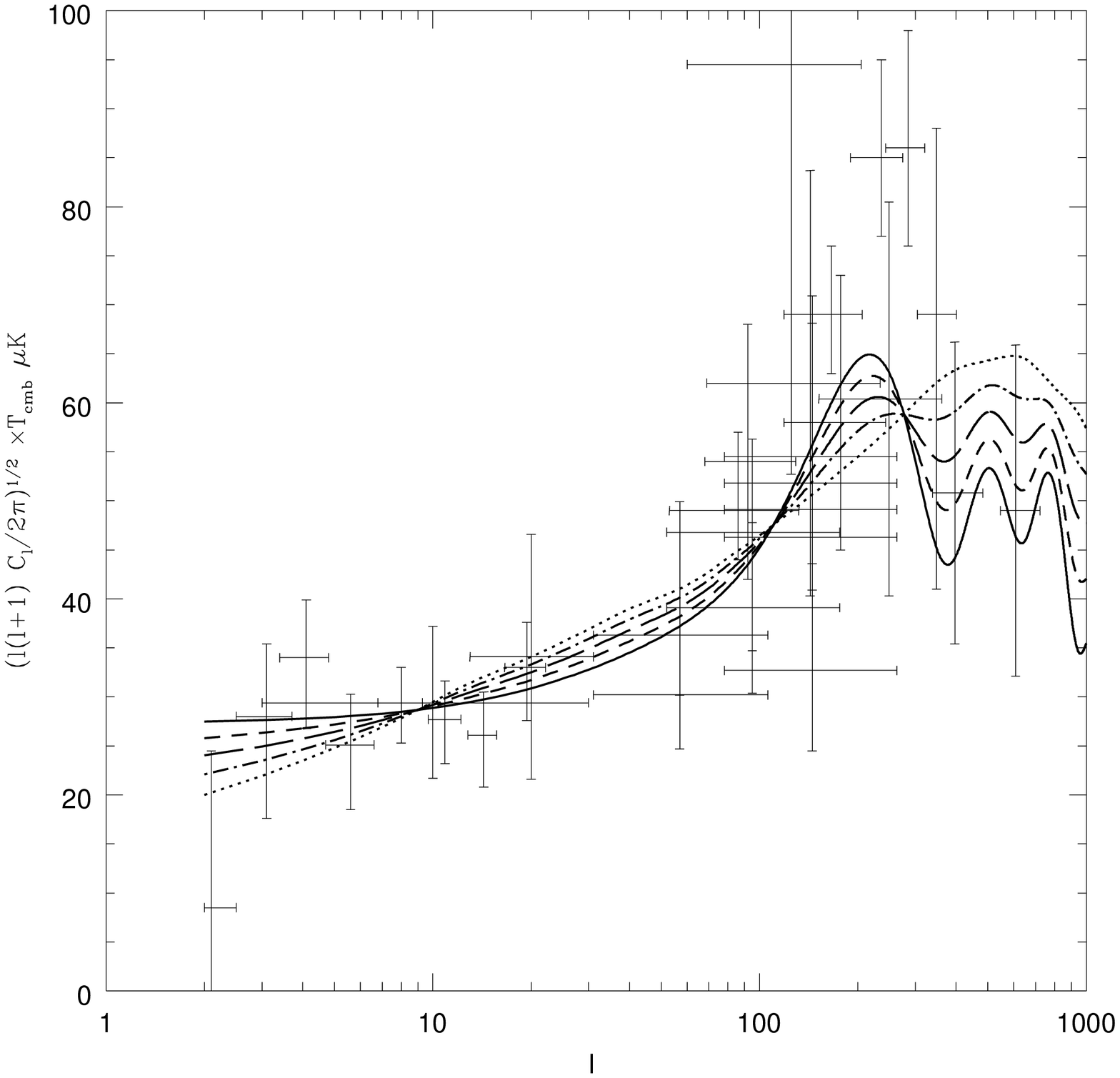,width=3.4in}}
\end{minipage}\hfill
\begin{minipage}{8.0cm}
\rightline{\psfig{file=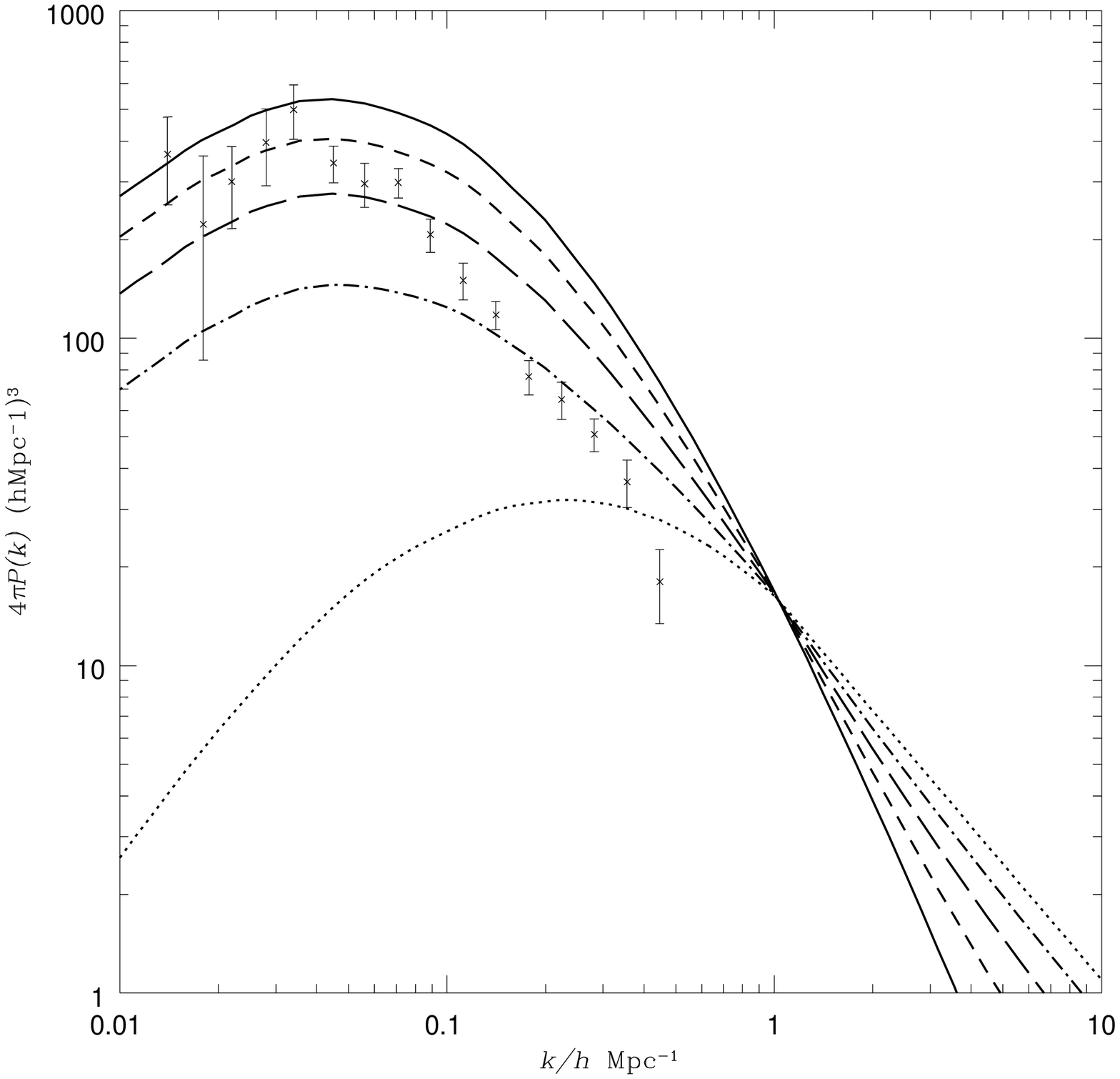,width=3.4in}}
\end{minipage}
\caption{On the left the angular power spectrum of CMB anisotropies and on the
right the power spectrum of fluctuations in the CDM for the standard CDM
scenario mixed with standard string source using a ratio of $\alpha=1.0$ (solid
line), $\alpha=0.75$ (short dash line), $\alpha=0.5$ (long dash line),
$\alpha=0.25$ (short dash-dotted line) and $\alpha=0.0$ (dotted line).} 
\label{fig-scdmstring}
\end{figure}

As a simple illustration of how to use this prescription for combining the
spectra, we have computed the angular power spectrum of CMB anisotropies and
the power spectrum of the matter fluctuations in the CDM for the standard CDM
scenario combined with both the standard scaling source (results presented in
fig.~\ref{fig-scdmscaling}) and the standard string model
(fig.~\ref{fig-scdmstring}), for $\alpha=1.0,0.75,0.5,0.25$ and $0.0$. These
figures illustrate the generic qualitative behaviour that one might expect in
these mixed perturbation scenarios. 

The value of $\sigma_8$ can be  computed from $P(k)$ via the formula
\beq
\sigma_8^2=4\pi\int{dk\over k}k^3P(k)|W(8kh^{-1}{\rm Mpc})|^2\,,
\eeq
where the window function $W(x)$, given by $W(x)=3\left(\sin x - x\cos
x\right)/x^3$. Hence, its value
in these mixed perturbation scenarios is given by 
\beq
\sigma_8^2=\alpha\left(\sigma_8^{\rm
adia}\right)^2+(1-\alpha)\left(\sigma_8^{\rm str}\right)^2\,, 
\label{sigmaadiastr}
\eeq
where $\sigma_8^{\rm adia}$ and $\sigma_8^{\rm str}$ are the COBE normalized
values for the individual components. Therefore, we see that the values of
$\sigma_8$ add in quadrature weighted by the factors $\alpha$ and
$1-\alpha$. If one of the computed values for the individual components is
below the observed value and the other is above, then it is possible to choose
$\alpha$ so that the mixture gives the observed value, $\sigma_8\approx
0.6$. For the combination of standard CDM and the standard scaling source, we
find that the value of $\alpha$ which does this is $\alpha\approx 0.20$,
whereas for the standard string model $\alpha\approx 0.15$. Both these values
are low reflecting the fact that on small scales the strings dominate, and
hence on the larger scales where the strings appear to be deficient, the
appealing aspects of the adiabatic perturbations are lost. We shall see in the
next section that this is a robust feature of models using the standard
cosmological parameters. 

\subsection{General hybrid inflation models combined with strings}
\label{sec-genhybridstrings}

In this section we will discuss qualitatively the observational consequences of
allowing for a string induced component to the CMB anisotropies and the
fluctuations in the CDM, in addition to an adiabatic component which is assumed
to come from inflation. Our treatment is totally general, applying to any
inflationary scenario, but specifically we have in mind the GUT scale Hybrid
Inflation scenario discussed in section~\ref{sec-genhybrid}. We will allow for
the moment for the ratio of the two components to be arbitrary and for the
spectral index to vary in the range $0.7<n<1.3$. From the observational point
of view, we will start conservatively with simple tests which compare the
amplitude of fluctuations in the CMB and the matter distribution, before
discussing the shape of the observed power spectrum. Initially, we will
concentrate on a universe with critical matter density with a matter content
which comprizes only of CDM and baryons. We will see that even with the extra
string induced component under discussion here it is difficult to fit all the data without
relaxing either of these assumptions. 

\begin{figure}
\setlength{\unitlength}{1cm}
\begin{minipage}{8.0cm}
\leftline{\psfig{file=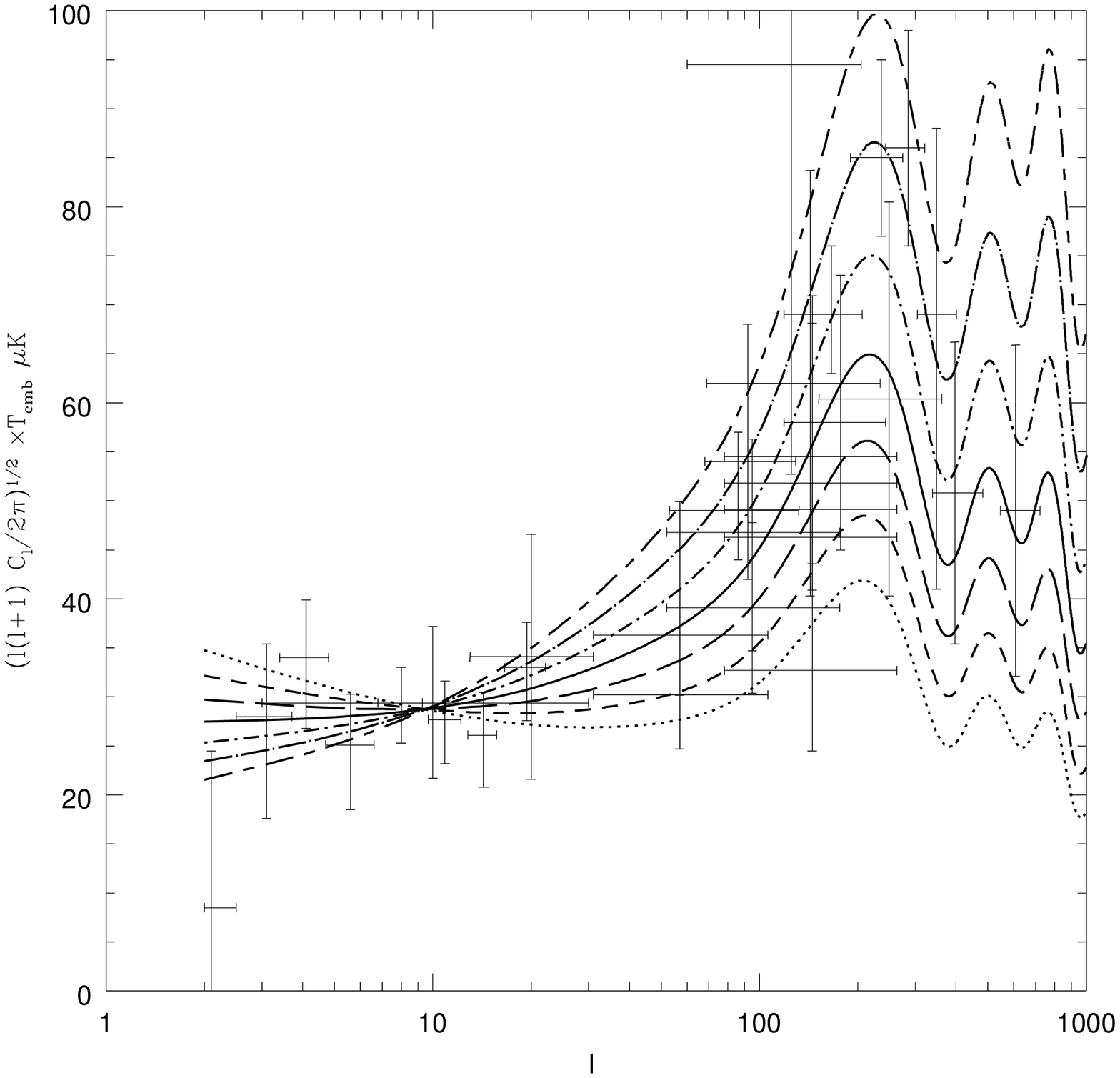,width=3.4in}}
\end{minipage}\hfill
\begin{minipage}{8.0cm}
\rightline{\psfig{file=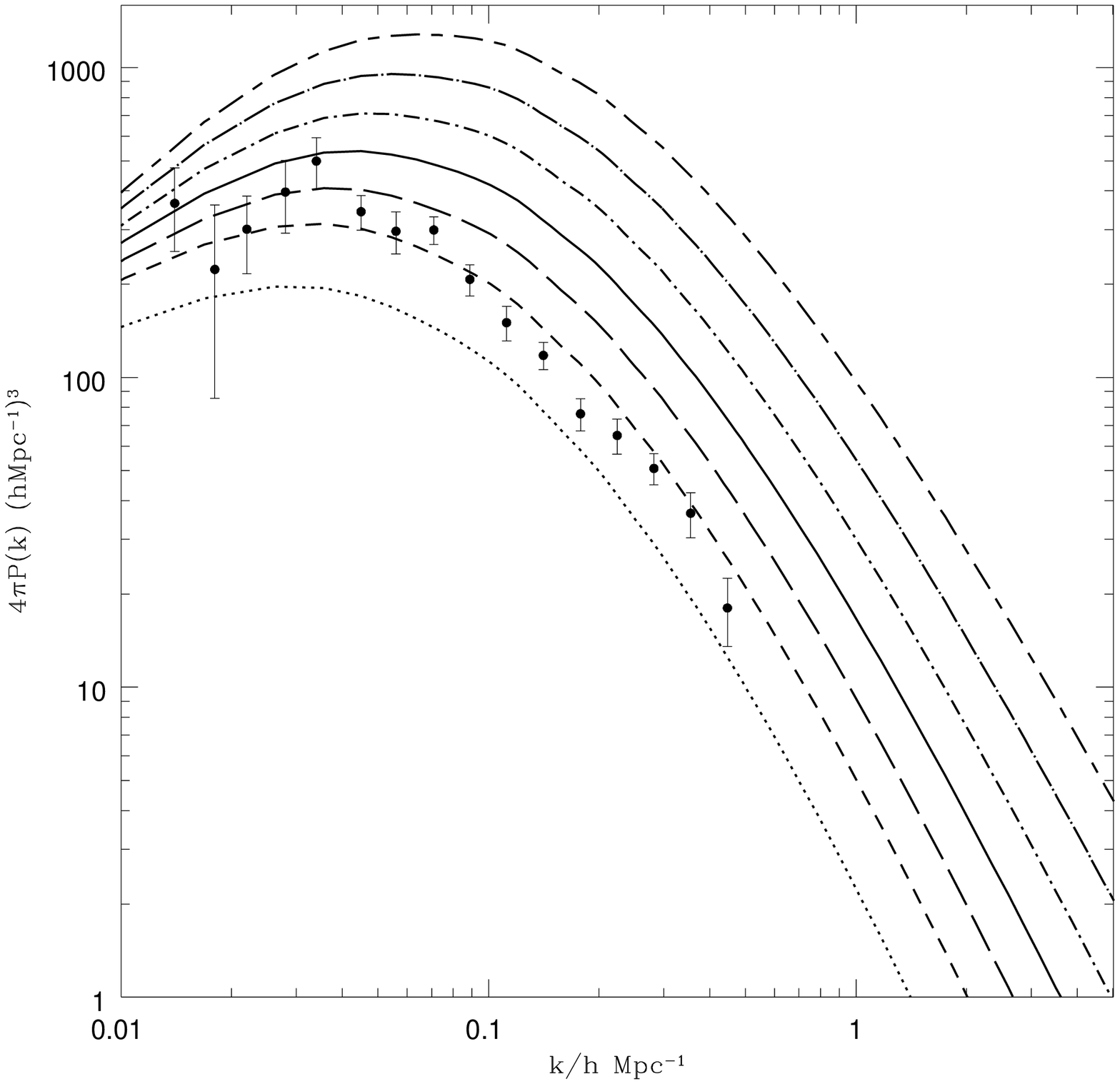,width=3.4in}}
\end{minipage}
\caption{On the left the angular power spectrum of CMB anisotropies and on the
right the power spectrum of the fluctuations in the CDM  for models with the
spectral index $n$ varying between 0.7 and 1.3. $n=0.7$ (dotted line), $n=0.8$
(short dash line), $n=0.9$ (long dash line), $n=1.0$ (solid line), $n=1.1$
(dot-short dash line), $n=1.2$ (dot-long dash line), $n=1.3$ (short dash-long
dash line). At this stage no string induced component has been included.} 
\label{fig-index}
\end{figure}

The COBE normalized spectra of the CMB and CDM for adiabatic component are
presented in fig.\ref{fig-index} for range of values of $n$, and clearly none of the models does particularly
well with respect to all the observations. The models with low $n\approx 0.8$
give a good fit to the observed matter power spectrum, assuming no bias, while
giving an apparently poor fit to the observations of the anisotropy in the CMB
on small angular scales. For larger values of $n\approx 1.2$, the situation is
reversed with the fit to the matter power spectrum requiring some kind of scale
dependent bias, while at least on smaller angular scales the comparison with
the measurements of the CMB is much better, although we note that the large
angle spectrum is only marginally compatible with the spectrum of anisotropies
detected by COBE. It is this rather unsatisfactory situation, which leads
joint analyses of the two different types of measurements to conclude that the
best fit to the data is given by something close to $n=1$. 

Probably the most stringent and robust constraint on any 
model for structure formation comes from comparing the magnitude of the CMB anisotropies detected by
COBE with the amplitude of the measured matter fluctuations on $8h^{-1}{\rm
Mpc}$, $\sigma_8$. Assuming that we can estimate the observed value of
$\sigma_8$ in the underlying matter distribution, without recourse to bias or
anti-biasing, then it is possible to rule out a large class of models. In
particular, as we have already discussed this test would rule out the standard
CDM scenario, unless an exotic anti-biasing mechanism was at work. 

\begin{figure}
\setlength{\unitlength}{1cm}
\begin{minipage}{8.0cm}
\leftline{\psfig{file=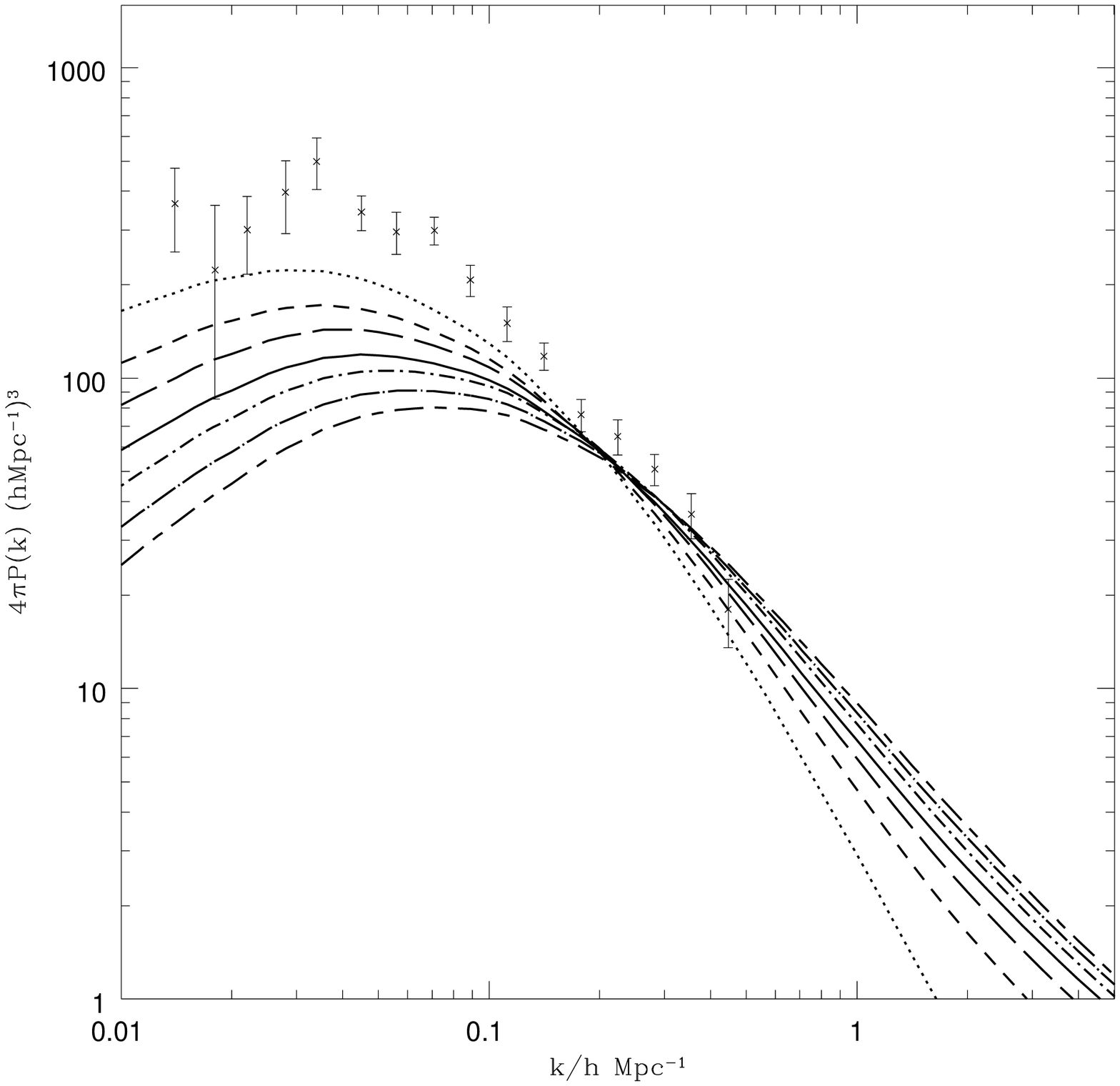,width=3.4in}}
\end{minipage}\hfill
\begin{minipage}{8.0cm}
\rightline{\psfig{file=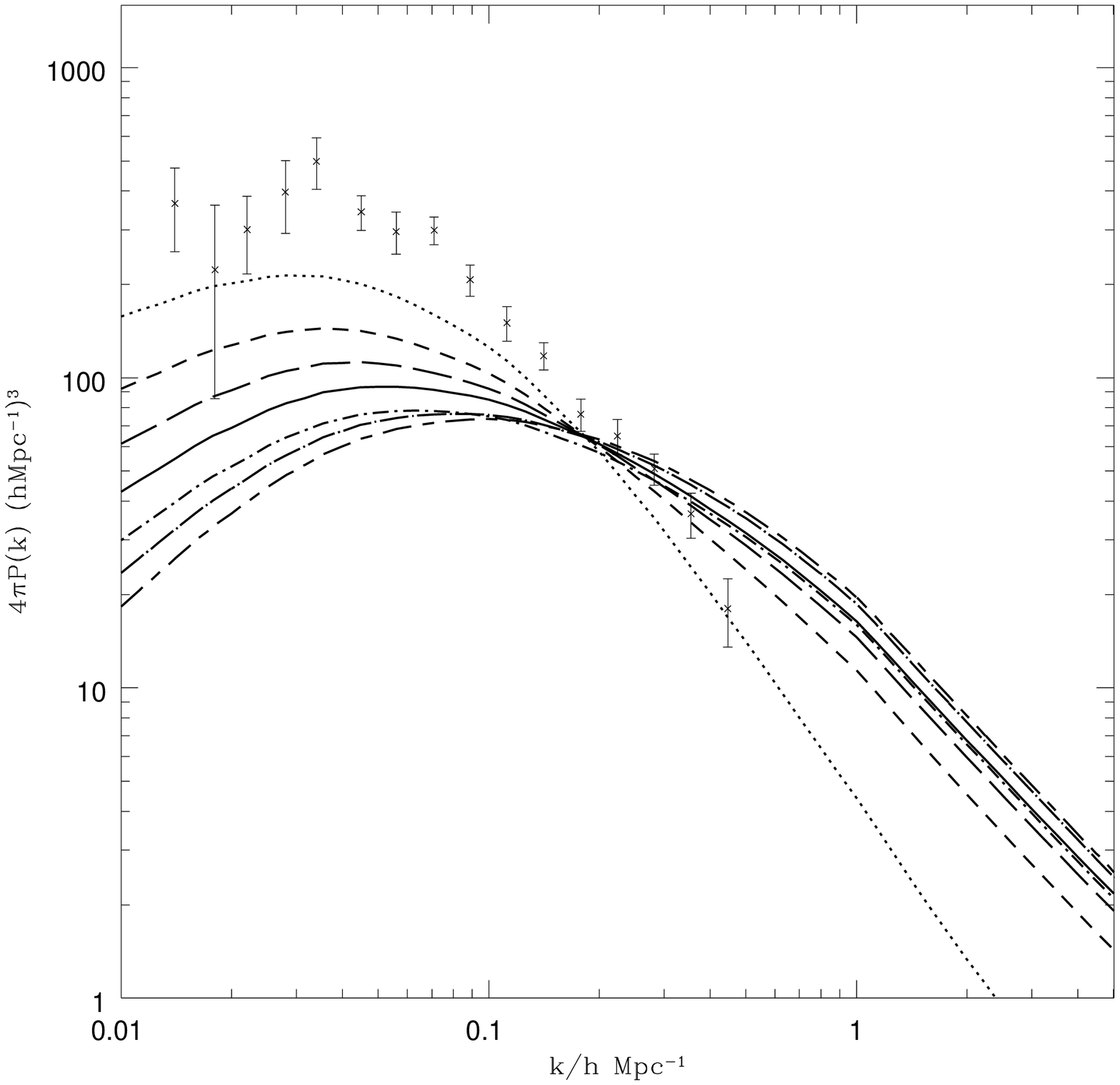,width=3.4in}}
\end{minipage}
\caption{Mixed perturbation scenarios which require no bias to fit the observed
value of $\sigma_8$ with $n$ varying between 0.7 and 1.3 for the adiabatic
component. On the left the defect component is that of the standard scaling source and on the right it is that of the standard string source. The curves are labelled as in fig.~\ref{fig-index}} 
\label{fig-alphaindex}
\end{figure}

We will assume, from the point of view of this exercise, that the observed
value of $\sigma_8=0.6$ and we will attempt to construct COBE normalized models
which can fit this value. This can be done simply by computing $\alpha$ using
(\ref{sigmaadiastr}) for given values of $\sigma_8^{\rm adia}$ and
$\sigma_8^{\rm str}$. The computed values of $\alpha$ are given in
table~\ref{tab-spect} for both the standard scaling and string scenarios, and
the resulting CDM power spectra are presented in fig.\ref{fig-alphaindex}. We
see that, in both cases, the spectrum is dominated by the adiabatic component
on large scales and by the string induced component on small scales, with the
transition taking place around $k\approx 0.2h{\rm Mpc}^{-1}$. While the
amplitude of $\sigma_8$ is fitted exactly, the shape of mixed power spectra does not correspond to that which is observed. One could argue that the
observations on large scales are much less certain than the amplitude of
$\sigma_8$ and that such models will only be ruled out once more accurate data
is available on scales greater than around $20h^{-1}{\rm Mpc}$. We should note
that there is an interesting qualitative  difference between using the standard
scaling source and the standard string source on small scales, we shall discuss
this phenomenon further in section~\ref{sec-smallscales}. 

\begin{table}
\begin{center}
\begin{minipage}{8.0cm}
\begin{tabular}{cccc}
%\hline 
$n$ & $\sigma_8^{\rm adia}$ & $\alpha_1$ & $\alpha_2$ \\
\hline 
0.7 & 0.62 & 0.92 & 0.88\\
0.8 & 0.77 & 0.53 & 0.44\\
0.9 & 0.95 & 0.33 & 0.25\\
1.0 & 1.18 & 0.20 & 0.15\\
1.1 & 1.47 & 0.13 & 0.09\\
1.2 & 1.82 & 0.08 & 0.06\\
1.3 & 2.25 & 0.05 & 0.04\\
%\hline
\end{tabular}
\end{minipage}
\end{center}
\medskip
\caption{The computed values of $\sigma_8$ for pure adiabatic models using the standard
cosmological parameters, and a varying spectral index $n$. Included also are
the values of the ratio, $\alpha$, of the adiabatic and string induced
components, if a such a model is to give the observed value of $\sigma_8\approx
0.6$ in a critical density universe. The value $\alpha_1$ is the ratio when the
string induced component is that of the standard scaling source, that is,
$\sigma_8^{\rm str}\approx 0.31$, and $\alpha_2$ is that for the standard
string source, that is, $\sigma_8^{\rm str}\approx 0.42$.} 
\label{tab-spect}
\end{table}

\begin{figure}
\setlength{\unitlength}{1cm}
\begin{minipage}{8.0cm}
\leftline{\psfig{file=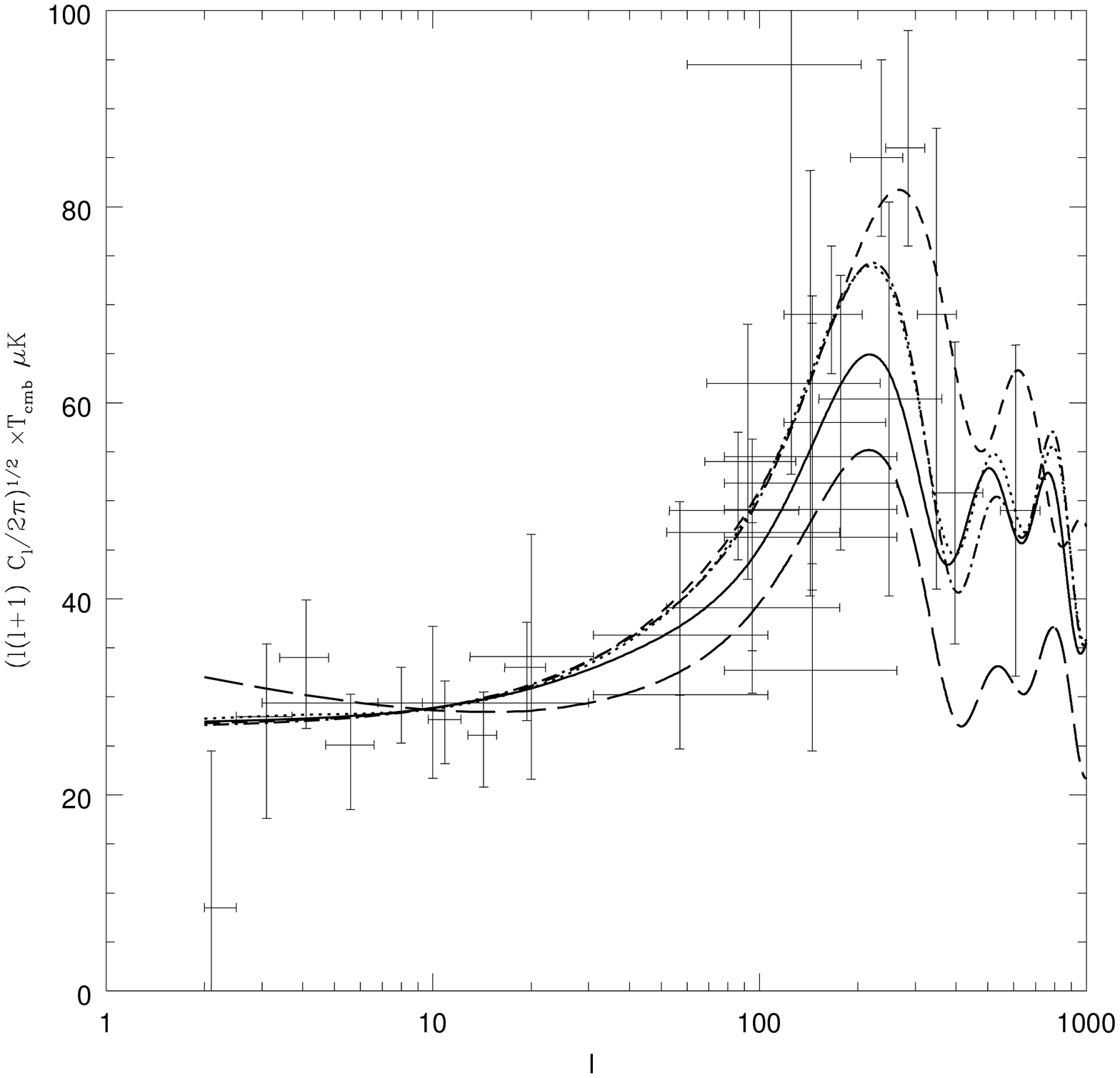,width=3.4in}}
\end{minipage}\hfill
\begin{minipage}{8.0cm}
\rightline{\psfig{file=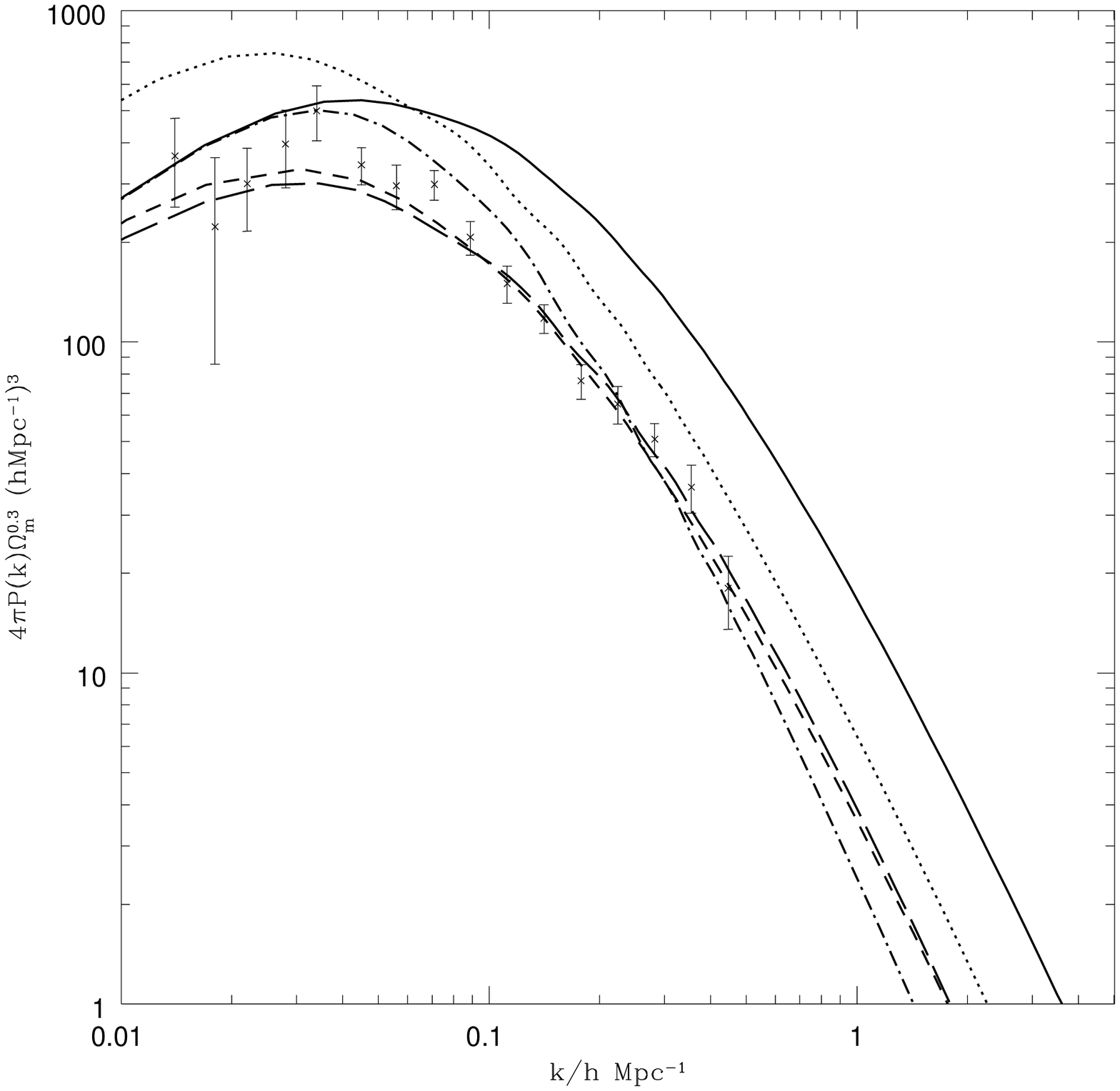,width=3.4in}}
\end{minipage}
\caption{On the left the angular power spectrum of CMB anisotropies and on the
right the power spectrum of fluctuations in the CDM for the best fit CDM type
models whose parameters are tabulated in table \ref{tab-models}. These models are designed to fit the shape of the observed power spectrum, but note that some of the models (B and D) require anti-biasing to fit the amplitude. (A,solid line) SCDM,
(B,dot-short dash line) CHDM, (C,long dash line) TCDM, (D,dotted line)
$\Lambda$CDM and (E,short dash line) hCDM. In both cases the current
observational data points are also included to guide the eye.} 
\label{fig-best}
\end{figure}

It appears that it is not possible to fit the exact shape of the observed power
spectrum by just varying the spectral index in a universe with the standard
cosmological parameters. A better fit to the current observations may be
achieved by varying the cosmological parameters namely the Hubble constant,
$h$, the number of massive neutrinos, $N_{\nu}$, and the contributions to the
cosmological density from CDM, $\Omega_{\rm c}$, Hot Dark Matter (HDM) such as
neutrinos, $\Omega_{\nu}$, the baryonic matter, $\Omega_{\rm b}$, and the
vacuum energy in the form of a cosmological constant, $\Omega_{\Lambda}$. A
recent analysis of pure adiabatic models~\cite{GawSilk} varying these
parameters suggests that the models whose parameters are tabulated in table~\ref{tab-models}
along with the computed values of $\sigma_8$ (included also is the Standard
Cold Dark Matter scenario) give the best fit to the current observations. The
anisotropies in the CMB and the fluctuations in the CDM are plotted in
fig.~\ref{fig-best}. Note that models B, C, D and E all fit the shape of the
observed matter power spectrum very much in contrast to model A, but that in
models B and D anti-biasing, that is, a bias of less than one (in fact, $b_{\rm
CHDM}\approx 0.85$ and $b_{\Lambda{\rm CDM}}\approx 0.7$), is required to
reconcile the amplitude of the spectrum with that of the observations (in
refs.~\cite{PDa,PDb}, the bias of IRAS galaxies was assumed to be one which means that they are good tracers of the underlying mass distribution. This may not necessarily be true). 

\begin{figure}
\setlength{\unitlength}{1cm}
\centerline{\psfig{file=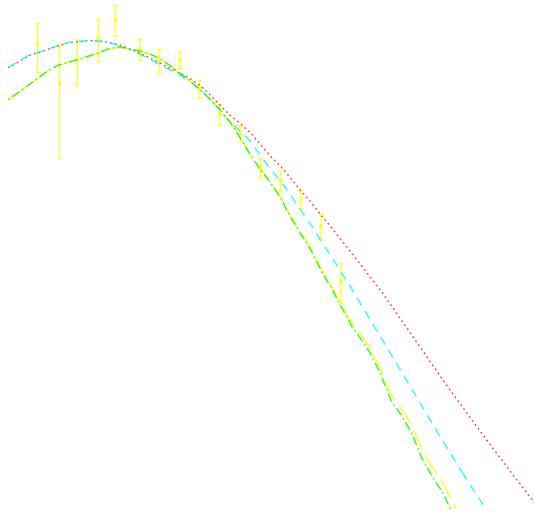,width=3.4in}}
\caption{The power spectrum of the fluctuations in the CDM for models B and D,
with a string induced component computed using the standard scaling source (dot-short dash line and short dash line respectively) and the standard string source (long-dash line and dotted line respectively). For model B, $\alpha=0.7$ --- $70\%$ adiabatic
fluctuations  and $30\%$ from strings, while for model D, $\alpha=0.5$ ---
equal proportions of adiabatic and string induced fluctuations.  It is clear
that each of these models fits the observations very well in the linear regime,
$k<0.2h{\rm Mpc}^{-1}$, without the need for bias.} 
\label{fig-comb}
\end{figure}

We have already discussed the inclusion of a string induced component to model
A, and models C and E appear to fit the data extremely well without any
modifications, but the anti-biasing required for models B and D to fit the data
could be perceived as a problem for such scenarios. However, the inclusion of a
defect induced component (from either the standard scaling source or the standard string source) with the correct amplitude, $\alpha=0.7$ for model B
and $\alpha=0.5$ for model D, has exactly the desired effect on large scales,
as shown in fig.~\ref{fig-comb} 

\begin{table}
\begin{center}
\begin{tabular}{cccccccccc}
%\hline 
Model & Description & $\Omega_{\rm c}$ & $\Omega_{\rm b}$ & $\Omega_{\nu}$ &
$\Omega_{\Lambda}$ & $h$ & $n$ & $N_{\nu}$ & $\sigma_8$ \\ 
\hline 
A & SCDM & 0.95 & 0.05 & 0.00 & 0.00 & 0.5 & 1.0 & 0 & 1.18\\
B & CHDM & 0.70 & 0.10 & 0.20 & 0.00 & 0.5 & 1.0 & 1 & 0.83\\
C & TCDM & 0.90 & 0.10 & 0.00 & 0.00 & 0.5 & 0.8 & 0 & 0.69\\
D & $\Lambda$CDM & 0.45 & 0.05 & 0.00 & 0.50 & 0.6 & 1.0 & 0 & 1.03\\
E & hCDM & 0.95 &  0.05 & 0.00 & 0.00 & 0.3 & 1.0 & 0 & 0.68\\
%\hline
\end{tabular}
\end{center}
\medskip
\caption{The cosmological parameters of the models whose CMB anisotropies and
CDM fluctuations are plotted in fig.~\ref{fig-best}.} 
\label{tab-models}
\end{table}

In summary, therefore, the introduction of this extra degree of freedom allows
us to fix exactly the amplitude of $\sigma_8$, relaxing any constraint on $n$
from the simple test of comparing the amplitude of CMB anisotropies with $\sigma_8$, but this is at the expense of having too little power on
large scales. In fact, if one only uses this simple test, there is probably
very little constraint on the initial power spectrum on large scales, since the
strings can account for the CMB anisotropies observed by COBE. However, if one allows
for  a small component of HDM or a non-zero cosmological constant, then it is
possible to fit the all the observational data without the need to postulate
any kind of bias between the observations and the computed CDM power
spectrum. It is also worth noting that each of these mixed scenarios does much
better on large scales than the string induced spectrum by itself, mildly
relieving the  so called `$b_{100}$ problem'~\cite{ABRa,ABRb}. We will discuss
the important features of the CMB power spectrum induced in these models in section~\ref{sec-sigcmb}.  
 
\subsection{Observational aspects of the Linde-Riotto model}
\label{sec-super}

\begin{figure}
\setlength{\unitlength}{1cm}
\centerline{\psfig{file=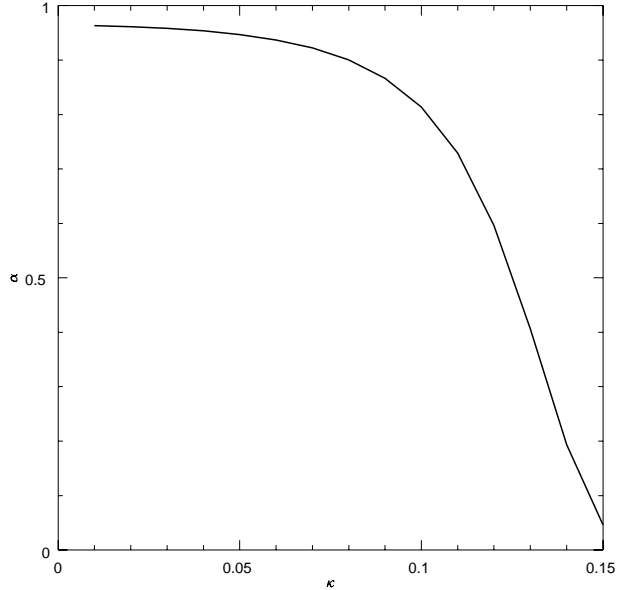,width=3.4in}}
\caption{The relative normalization of the adiabatic and string induced
components, $\alpha$, in the Linde and Riotto model plotted as a function of
the model parameter $\kappa$. The case of  $\alpha\approx 1$ corresponds to most of the fluctuations being adiabatic, and $\alpha\approx 0$ corresponds
to most of them being induced by strings.} 
\label{fig-alpha}
\end{figure}

We will now turn our attention to the specific Supergravity-inspired model of
inflation discussed in section~\ref{sec-linrio}. There, it was shown that for
$\kappa$ small the predictions were very similar to that of a model with
scale-free spectrum, but that for larger values of $\kappa<\kappa_{\rm c}$ more
exotic initial spectra were possible. We also noted that for generic symmetry
breaking schemes the model would lead to the production of cosmic strings whose
mass per unit length would be large enough for them to have a substantial
effect on cosmic structure formation. If one considers these models as
candidates for the mixed perturbation scenarios parameterized by the relative
normalization $\alpha$, then the normalization of the strings will be given by

\beq
{G\mu\over 1\times 10^{-6}}\approx \sqrt{1-\alpha}
\label{gmunorm}
\eeq
and the normalization of the adiabatic perturbations requires that
\beq 
1.7\times
10^{-5}\sqrt{\alpha}\approx\displaystyle{\gamma^2\over\kappa^{3/2}}\left[\sin\left({30\kappa\over\pi}\right)\right]^{1/2}\left[\cos\left({30\kappa\over\pi}\right)\right]^{3/2}\,.
\eeq
The value of $G\mu$ can be computed in terms of $\gamma$ and $\kappa$ and
therefore one can use (\ref{gmunorm}) to eliminate $\gamma$, to give $\alpha$
in terms of $\kappa$,  
\beq
\alpha^{-1}=1+0.36\kappa\left[\sin\left(\displaystyle{30\kappa\over\pi}\right)
\cos^3\left({30\kappa\over\pi}\right)\right]^{-1}\,. 
\eeq
This function can be approximated in both the limit of $\kappa$ small, in which
case $\alpha\approx 1$, and $\kappa\approx\kappa_{\rm c}$ where $\alpha\approx
0$. In these two limiting cases one or the other of the two sources dominates, but in the more general case any relative normalization of the
two components is possible. The precise function $\alpha(\kappa)$ is plotted in
the range $0<\kappa<\kappa_{\rm c}$ in fig.\ref{fig-alpha} and it is tabulated
for various values of $\kappa$ in table~\ref{tab-sup} along with the values of
$G\mu$, $\gamma$ and $\sigma_8$ for the mixed scenario. Notice that all the
values of $\gamma$, which can be computed by using the energy units $E=M_{\rm
p}/\sqrt{8\pi}=2.4\times 10^{18}{\rm GeV}$, lie in the sensible range of
$1.9\times 10^{15}{\rm GeV}$ for $\kappa=0.08$ and $4.6\times 10^{16}{\rm GeV}$
for $\kappa=0.15$, and that the corresponding values of $G\mu$ are even more
favourable relative to the constraint from milli-second pulsars~\cite{CBS}. 

\begin{table}
\begin{center}
\begin{minipage}{8.0cm}
\begin{tabular}{ccccc}
\hline 
$\kappa$ & $\alpha$ & $G\mu$ & $\gamma$ & $\sigma_8$\\
\hline 
0.08 & 0.90 & $3.2\times 10^{-7}$ & $8.0\times 10^{-4}$ & 1.20\\
0.10 & 0.81 & $4.4\times 10^{-7}$ & $1.1\times 10^{-3}$ & 1.25\\
0.12 & 0.60 & $6.3\times 10^{-7}$ & $1.4\times 10^{-3}$ & 1.28\\
0.13 & 0.41 & $7.7\times 10^{-7}$ & $1.6\times 10^{-3}$ & 1.23\\
0.14 & 0.19 & $9.0\times 10^{-7}$ & $1.8\times 10^{-3}$ & 1.11\\
0.15 & 0.05 & $9.7\times 10^{-7}$ & $1.9\times 10^{-3}$ & 0.99\\
 \hline
\end{tabular}
\end{minipage}
\end{center}
\medskip
\caption{The relative contribution from adiabatic and string induced
fluctuations, $\alpha$ as a function of $\kappa$ for the Linde-Riotto model. Also included are the
corresponding values of $G\mu$ and $\gamma$ and the value of
$\sigma_8$ using a mixed perturbation scenario with the computed value of
$\alpha$.} 
\label{tab-sup}
\end{table}

\begin{figure}
\setlength{\unitlength}{1cm}
\begin{minipage}{8.0cm}
\leftline{\psfig{file=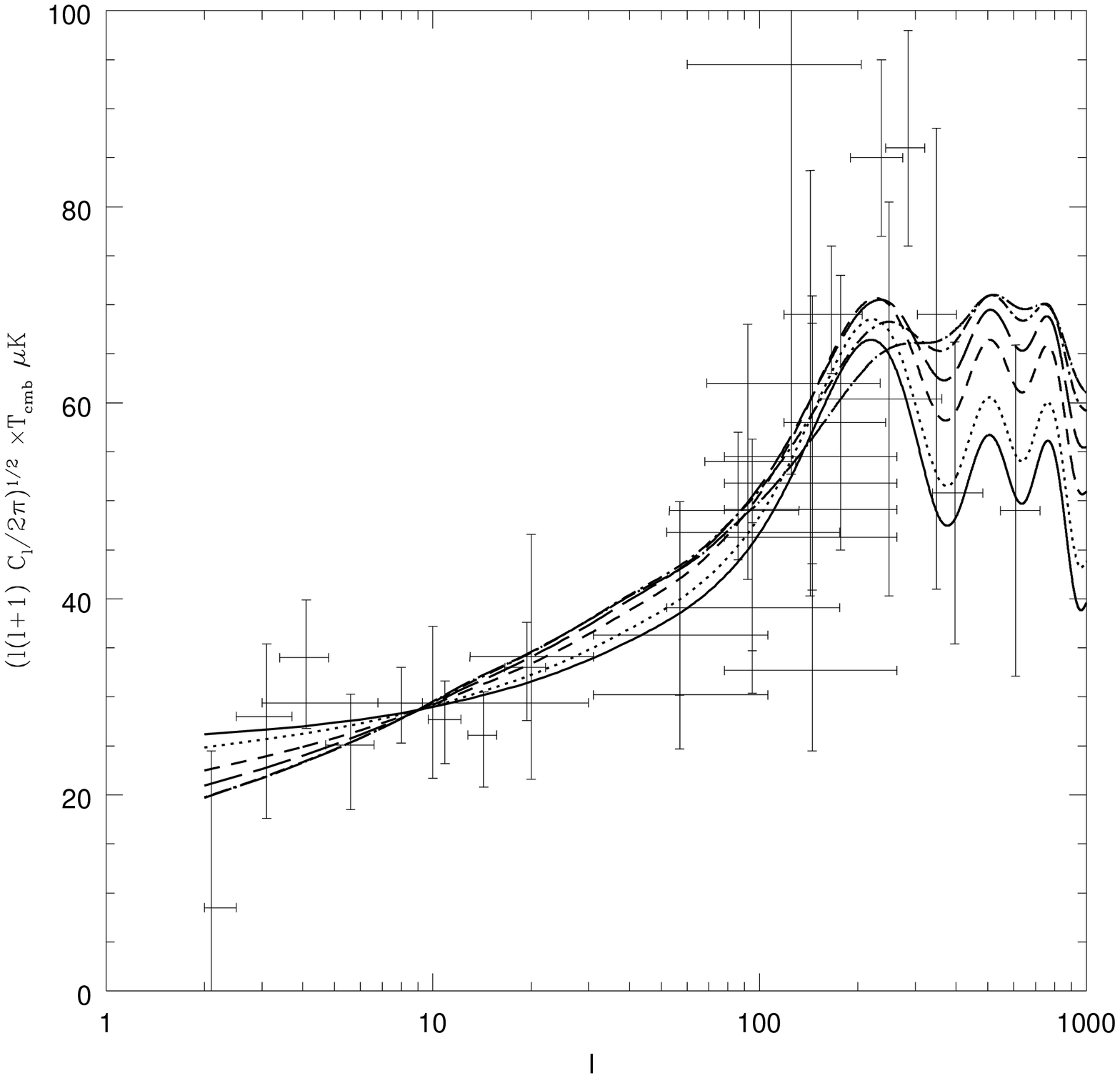,width=3.4in}}
\end{minipage}\hfill
\begin{minipage}{8.0cm}
\rightline{\psfig{file=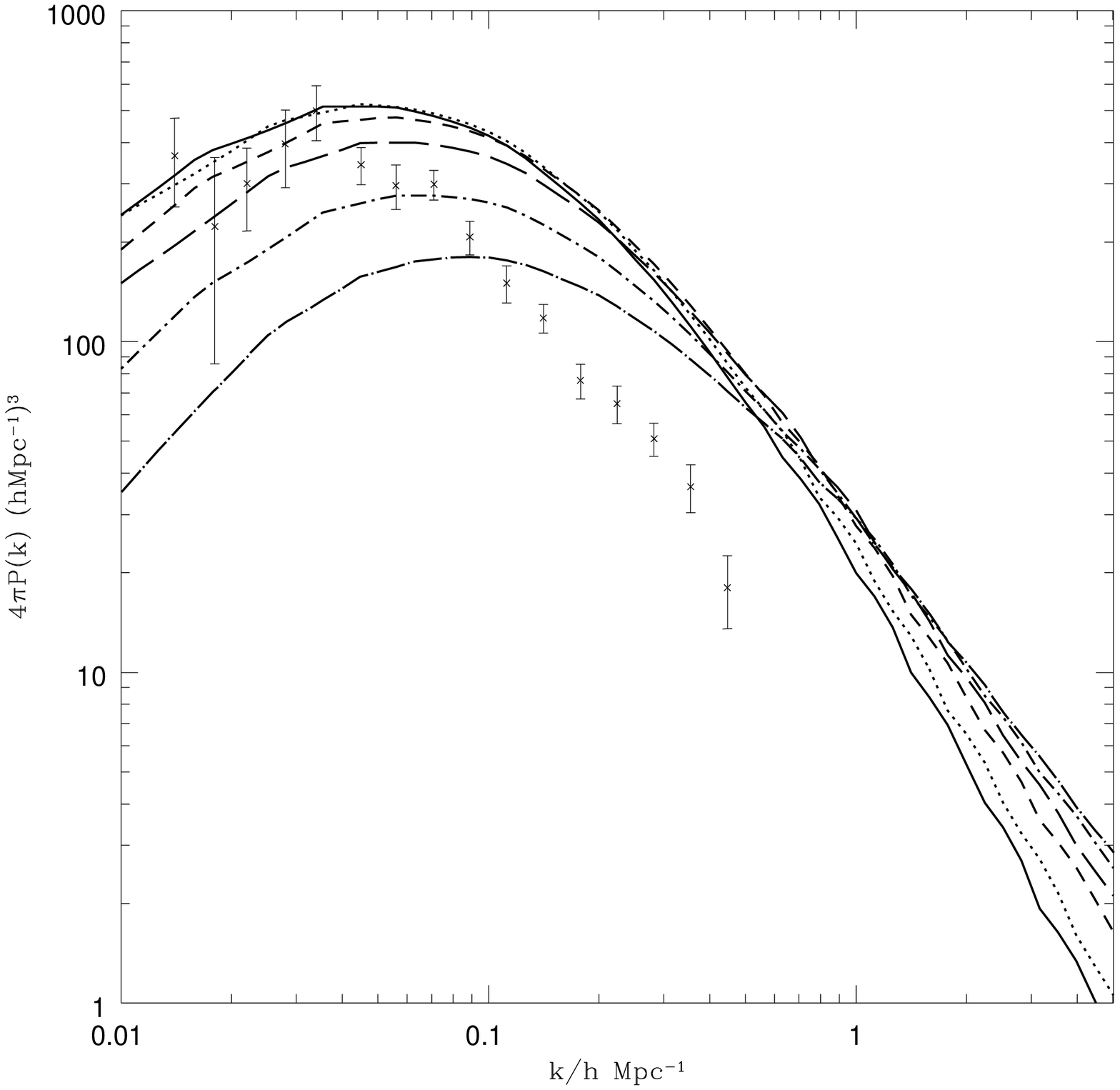,width=3.4in}}
\end{minipage}
\caption{The same quantities and models as in fig.~\ref{fig-varykappa} except
that we have included a string induced component (the standard string source
only) with the relative normalization given in table~\ref{tab-sup} for each
value of $\kappa$.} 
\label{fig-sugracdmstring}
\end{figure}

\begin{figure}
\setlength{\unitlength}{1cm}
\begin{minipage}{8.0cm}
\leftline{\psfig{file=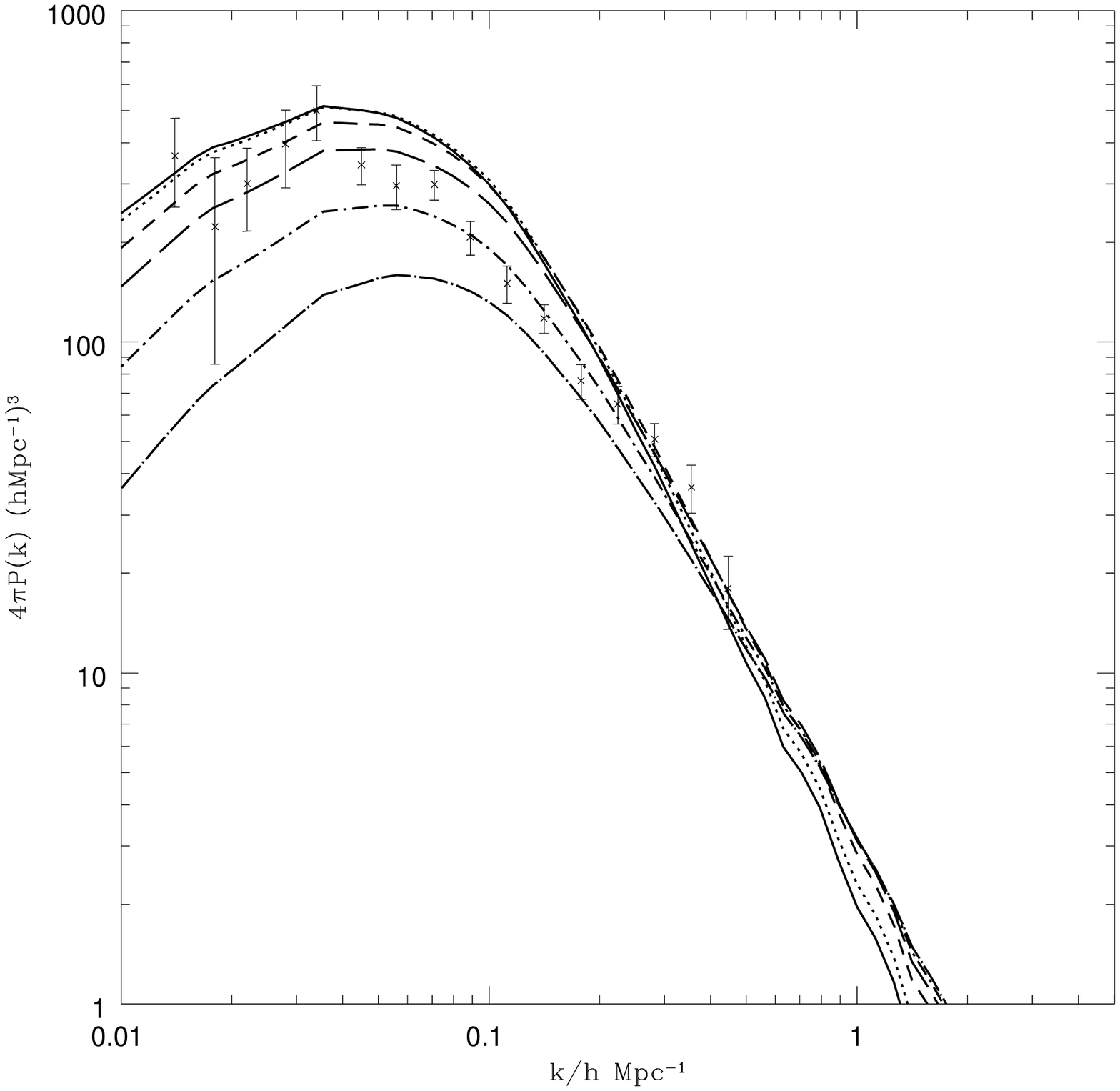,width=3.4in}}
\end{minipage}\hfill
\begin{minipage}{8.0cm}
\rightline{\psfig{file=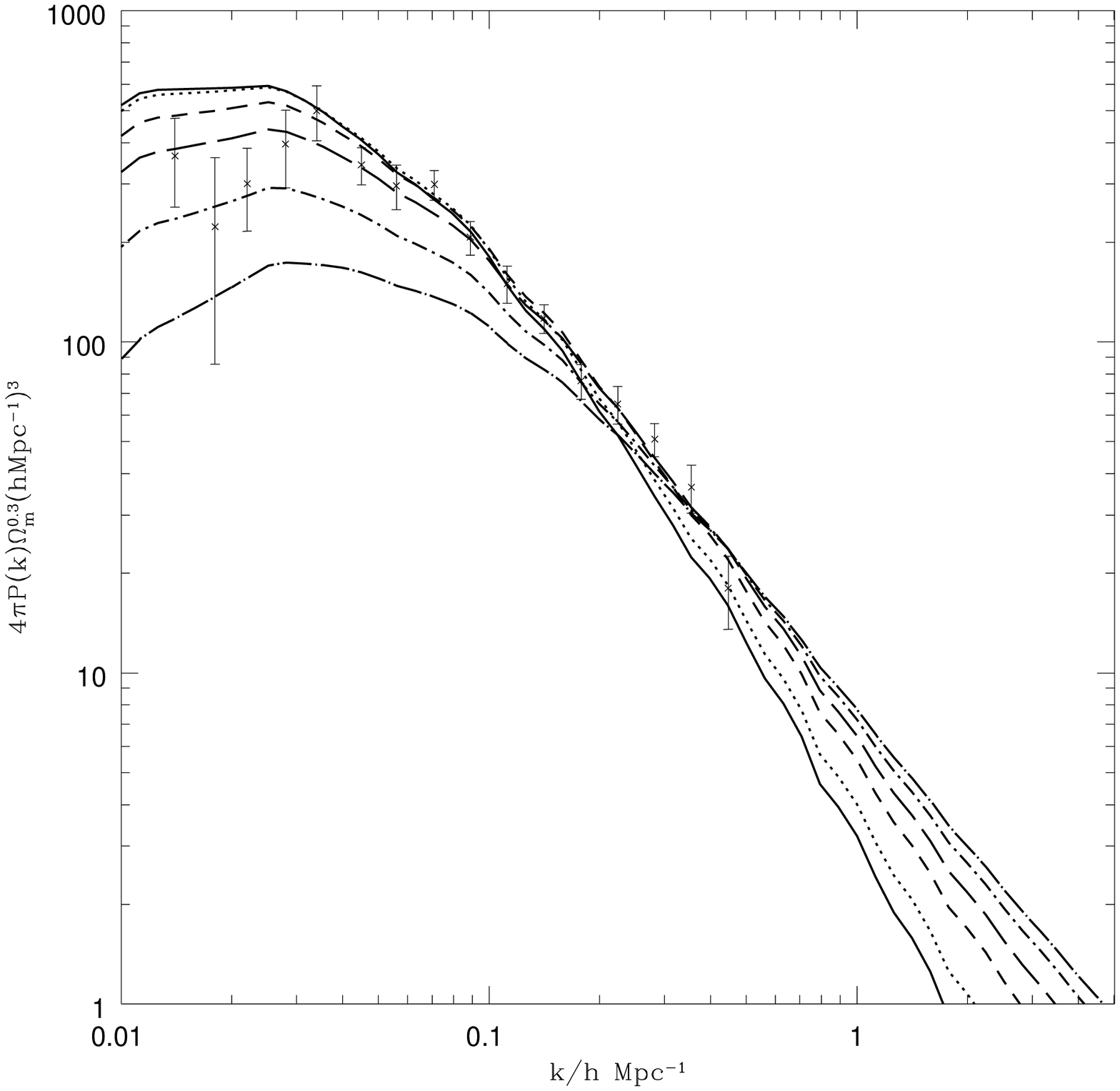,width=3.4in}}
\end{minipage}
\caption{On the left the CDM power spectrum for the Linde-Riotto model with a string induced component included for $\Omega_{\nu}=0.3$ and $N_{\nu}=1$, and on the right for $\Omega_{\Lambda}=0.6$. The curves are labelled as in figs.~\ref{fig-varykappa} and \ref{fig-sugracdmstring}. On the left $\kappa=0.14$ appears to give the best fit whereas on the right $\kappa=0.13$ is the best. Note that without the inclusion of the string induced component each of these models would be ruled out.}
\label{fig-lrdata}
\end{figure}

Using this relative normalization and the standard string source model for the defect induced component, we have computed the CMB anisotropies and
fluctuations in the CDM for the same values of $\kappa$ used in
fig.~\ref{fig-varykappa} and the properly normalized results are presented in
fig.~\ref{fig-sugracdmstring}. Notice that the models with large values of
$\kappa$ (for example, $\kappa=0.15$), which were wildly at odds with the
observations without the inclusion of the string induced components, appear to
have much more acceptable spectra, with all the computed values of $\sigma_8$
being around 1. Of course the shape of the spectrum is not quite correct, a
feature which is common to most sensible critical density CDM models ---
although see models C and E from section~\ref{sec-genhybridstrings} --- but it
is possible to rectify this situation by the inclusion of a small HDM component
or a non-zero cosmological constant. Fig.~\ref{fig-lrdata} shows the same
models using the standard cosmological parameters except that in one we have
replaced some of the CDM with HDM ($\Omega_{\nu}=0.3$ and $N_{\nu}=1$) and in
the other it has been replaced by a non-zero cosmological constant
($\Omega_{\Lambda}$=0.6). These are the kind of modifications to the
cosmological parameters which are well known to achieve a better fit to the
data. However, in the particular case of this inflationary model which
generically induces a blue initial spectrum, the amount of HDM or cosmological
constant required to achieve a good fit is slightly larger than in the scale
free case. We find that the best fit for these models is achieved for the CHDM
model with $\kappa=0.14$ and for the $\Lambda$CDM it is $\kappa=0.13$. It is
interesting to note that, without the string induced component, these models
would clearly be at odds with the observations, requiring substantial
anti-biasing to be compatible with the observed matter power spectrum and not
having anything close to a flat CMB power spectrum on small scales. 

\subsection{Novel features in matter power spectrum on small scales}
\label{sec-smallscales}

So far the discussion of these mixed perturbation scenarios has focussed on the linear part of the power spectrum ($k<0.2h{\rm Mpc}^{-1}$) and the fixing the value of $\sigma_8$. In this section, we will focus our attention on the behaviour of the power spectrum on small scales when a defect component is included. Investigating these small scales is fraught with complications since the power spectrum will be affected by non-linearity, but methods exist to use the linear spectra that we have computed in the previous sections without resorting to numerical N-Body simulations. These methods have been extensively tested using numerical codes for adiabatic models and we shall assume that one can also use them in the case of active fluctuations created by defect networks.

The basic feature that we will use is that the linear power spectra created by defects has much more small scale power than an adiabatic model with the same cosmological parameters. The basic reason for this being that the power spectrum of the strings falls off very slowly inside the horizon, that is, $\langle\Theta_{00}(k,\eta)\Theta^{*}_{00}(k,\eta)\rangle\propto k^{-2}$, whereas the adiabatic perturbations are created at horizon crossing with a sharp tail. Hence, there will be a feature in the power spectrum at some scale, which is marks the transition from the spectrum being dominated by adiabatic perturbations to string induced perturbations being dominant. Schematically it can be thought of as a kink in the spectrum, although as one can see from all the figures presented to date for these mixed scenarios it is often difficult to see it clearly with the naked eye. Formally, this corresponds to a change in the fall off of the spectrum on small scales.
 
\begin{figure}
\setlength{\unitlength}{1cm}
\begin{minipage}{8.0cm}
\leftline{\psfig{file=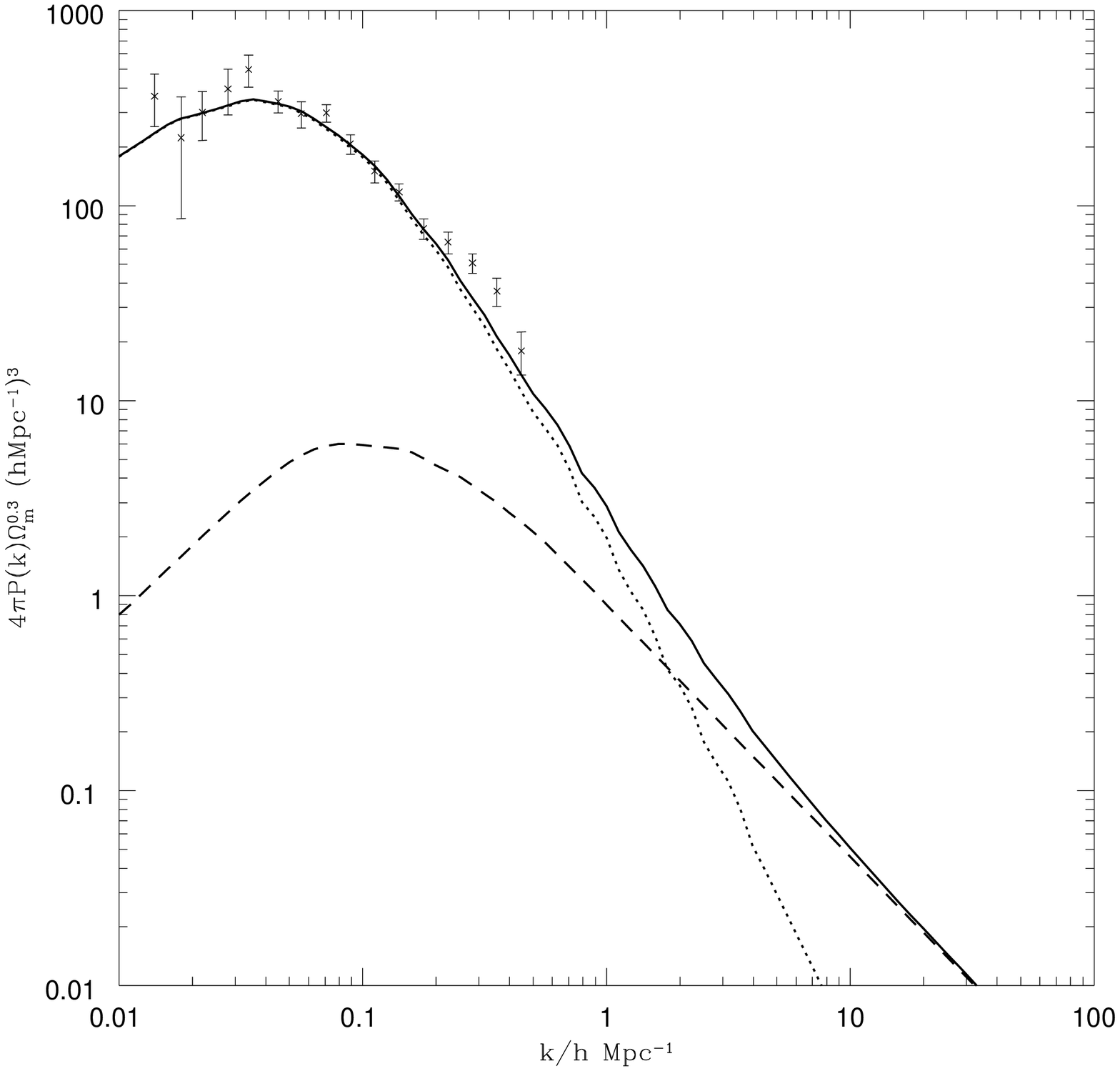,width=3.4in}}
\end{minipage}\hfill
\begin{minipage}{8.0cm}
\rightline{\psfig{file=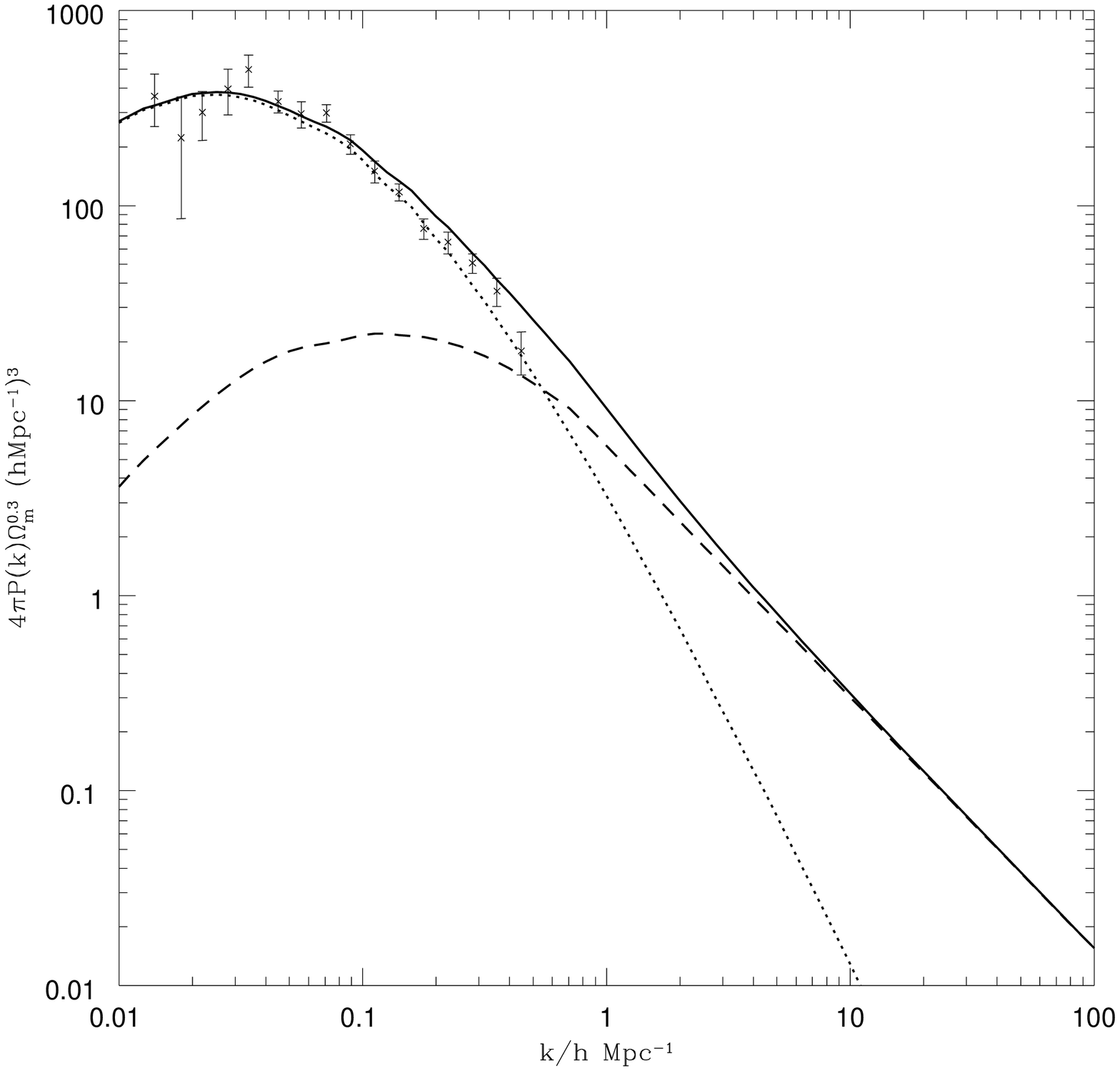,width=3.4in}}
\end{minipage}
\caption{The same spectra as in fig.~\ref{fig-comb}. On the left model B with a string induced component and relative normalization of $\alpha=0.7$, and on the right model D with a relative normalization of $\alpha=0.5$. The spectra are presented over a much wider range of wavenumbers than before, and the solid line corresponds to the mixed scenario with the dotted line just the adiabatic component and the dashed line that which is induced by strings. In the case of model D the mixed spectrum deviates significantly from the adiabatic one around $k\approx 0.5h{\rm Mpc}^{-1}$ and even around $k\approx 0.1h{\rm Mpc}^{-1}$ there are noticeable effects. Whereas for model B significant deviation from the adiabatic spectrum occur on smaller scales, around $k\approx 2h{\rm Mpc}^{-1}$.}
\label{fig-gensmallscale}
\end{figure}

We will just illustrate this effect first using the general models which were shown to fit the linear data well section~\ref{sec-genhybridstrings}, although this feature also manifests itself in the context of the Linde-Riotto model in almost exactly the same way. Fig.~\ref{fig-gensmallscale} shows models B and D with $\alpha=0.7$ and $\alpha=0.5$ respectively, as in fig.~\ref{fig-comb}, but with the adiabatic component and string induced component superposed. Clearly, the mixed spectra deviate significantly from the adiabatic ones on very small scales $k\approx 1h{\rm Mpc}^{-1}$, although in the case of model D there is a even noticeable difference on much larger scales around $k\approx 0.1h{\rm Mpc}^{-1}$. 

Recently, there have been two pieces of observational evidence which might plausibly point to features similar to these in the linear power spectrum. Although it it not totally clear whether these features are artifacts of analysis techniques, it is interesting to broaden the theoretical possibilities under consideration. Firstly, it has been reported that there exists just such a feature in the observed power spectrum~\cite{Pa,Pb,SKGPH}, once the effects of non-linearity have been removed. In these works the authors attempted to explain this feature as being due to complex biasing processes, since no simple CDM variant model appeared to be able to fit the data with just a linear bias. However, here we see that such a feature naturally occurs in these mixed perturbation scenarios which require no bias at all to agree with the observations. We are again assuming that the correct bias of the IRAS galaxies is one. Although this may not necessarily be the case, it does not effect our argument since at the moment we are allowing the relative normalization of the two components to be arbitrary, giving us the freedom to move the large-scale portion of the spectrum up and down. Of course, changing this relative normalization will modify the point at which the string induced component begins to dominate.

The second piece of observational evidence which might support these kind of features on small scales is the number of damped Lyman-$\alpha$ systems which are observed at high redshifts ($z\approx 4$). These measurements effectively correspond to an estimate of the same quantity as $\sigma_8$, but on much smaller scales. It has been shown that it is difficult to explain the observed amplitude  in the context of models such as CHDM (also in TCDM) which fit observations on the larger scales  since they
produce too little power on small-scales (see, for example, ref.~\cite{EH} and references therein). Conversely, models such as SCDM which appear to be at odds with the observations on large scales fair much better. Here, we see a simple modification to the model --- the inclusion of a defect induced component --- which creates more power on smaller scales. The extent to which this can improve the situation for these measurements will be discussed in another publication~\cite{BatW}, although it seems clear that given the freedom that we have in these scenarios, it should be possible to fit the data for at least some values of the parameters. It may be that the inclusion of a string induced component can account for the other observations of early structure formation in models which otherwise create too little power.

We should note that both these observational features rely heavily on our ability to continue the power spectrum into the non-linear regime. The comparison with the observations in this regime is much more complicated and less quantitative significance should attached to these points than the comparison with measurements in the non-linear regime. Nonetheless, they serve as an illustration of the kind of features which these mixed scenarios have. Assuming that the issue of bias can be understood, future redshift surveys (SDSS and 2Df) should be able to make accurate predictions for the power spectrum of the fluctuations in the CDM and hence it will be possible to shed more light on the possible existence of kinks in the power spectrum. In the meantime is seems sensible to investigate further any scenario which naturally has such features.

\subsection{Distinctive signatures in the CMB and ruling out mixed scenarios}
\label{sec-sigcmb}

In the previous sections we have concentrated on the consequences of these mixed perturbation scenarios for the formation of structure, only using the large-angle measurements of the CMB anisotropies to normalize the CDM power spectrum. While the measurements of the CDM power spectrum are relatively extensive from various redshift surveys, the comparison to the computed spectra is always clouded by issues such as bias. Therefore, more clean tests of the cosmological models are required and the measurements of the anisotropies in the CMB on small angular scales should provide more accurate data, free of systematic uncertainties such as bias. The amount of data amassing on smaller angular scales is already substantial and future satellite missions such as MAP and PLANCK should take the study of CMB anisotropies to a new dimension. Therefore, it seems sensible to discuss these possibilities, particularly since we will see that these mixed perturbation scenarios have a very distinctive signature.

The first thing to be aware of in this context is that the spectra for
adiabatic models and the active source models under consideration here are very
different. The adiabatic spectra generically have oscillations, whereas the
active spectra appear not to have these striking features, having just a single
rise. Clearly, the superposition of these spectra will have very distinctive
features dependent on which of the two components is dominant. The other
feature that is very different between the two different types of models is the
positions of the maxima in the spectrum. The SCDM model has its maxima around
$\ell\approx 200$, and all the other flat models have maxima around the same
place, but the standard scaling source model does not appear to have an obvious
peak in its spectrum and the standard string models has a peak at much smaller
scales around $\ell\approx 500$.   

\begin{figure}
\setlength{\unitlength}{1cm}
\begin{minipage}{8.0cm}
\leftline{\psfig{file=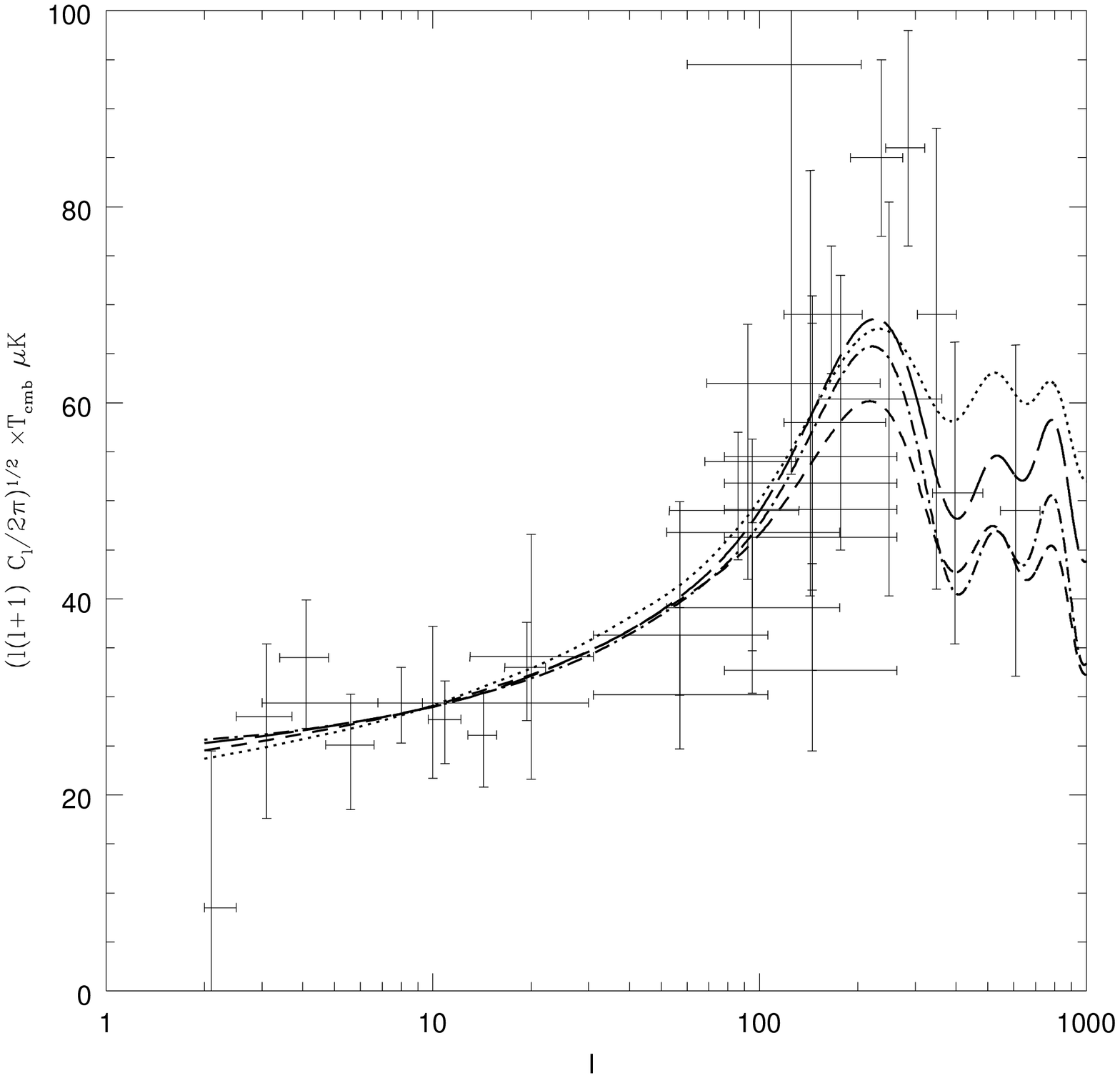,width=3.4in}}
\end{minipage}\hfill
\begin{minipage}{8.0cm}
\rightline{\psfig{file=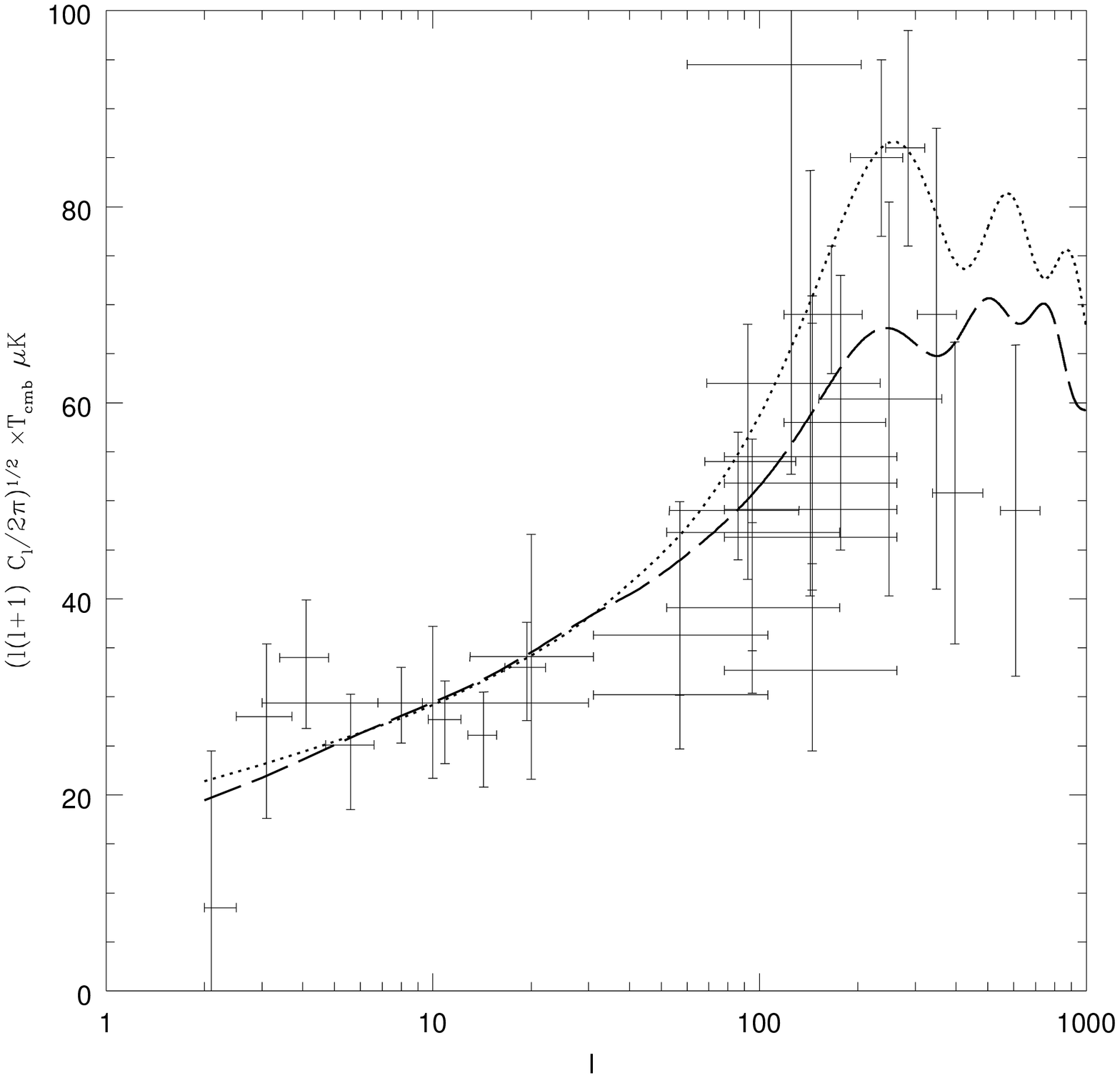,width=3.4in}}
\end{minipage}
\caption{On the left the CMB anisotropies for the models whose CDM power
spectra are presented in fig.~\ref{fig-comb}. The curves are labelled in the
same way as before. On the right are the CMB anisotropies for the Linde-Riotto
models which which fit the CDM power spectra well. The dotted line has
$\kappa=0.13$ and $\Omega_{\Lambda}=0.6$, and the long dash line has
$\kappa=0.14$, $\Omega_{\nu}=0.3$ and $N_{\nu}=1$. } 
\label{fig-cmbtemp}
\end{figure}

During the earlier discussions we have illustrated the effects of combining the
two types  of spectra for very simple cases and it is clear from this that a
wide variety of phenomena are possible. To illustrate some of the more
important points we have selected the models whose CDM power spectra are
presented in fig.~\ref{fig-comb} and the Linde-Riotto models which appear to
give the best fit to the observations of galaxy clustering by eye from
fig.~\ref{fig-lrdata}. The CMB anisotropies of these models are presented in
fig.~\ref{fig-cmbtemp}. 

In the general models, which have a scale free initial spectrum and an
arbitrary normalization of the two contributions, the spectra appear to be
dominated by the adiabatic fluctuations when the normalization is chosen to
give a bias of one relative to IRAS galaxies. When the defect induced component
is that of the standard scaling source then the strings only really contribute
on large scales,  bringing down the adiabatic component on small
scales. Whereas if one uses the standard string source, the opposite is true
and the smaller scales ($\ell>400$) are boosted relative to models with just
the adiabatic component. Of course the first adiabatic peak around $\ell\approx
200$ is always suppressed, since both the defect models are relatively low
there. It is also interesting to note the behaviour of the relative amplitudes
of the peaks since in a purely adiabatic model the modulation of the peaks due
to baryon drag has been suggested as a test of adiabatic
fluctuations~\cite{HWa,HWb}. Here, we see that the defect component acting as a
small background can change the relative sizes for a particular model, which
could create confusion when applying this test. 

In the models which give the best fit to the data for the Linde-Riotto
inflation model, that is, a CHDM model with $\Omega_{\nu}=0.3$, $N_{\nu}=1$ and
$\kappa=0.14$ and a $\Lambda$CDM model with $\Omega_{\Lambda}=0.6$ and
$\kappa=0.13$, the CMB anisotropies are dominated by the string induced
component on large scales and hence have a much smaller values of
$\alpha$. This leads to qualitatively different behaviour on smaller
scales. The oscillations in the adiabatic component now manifest themselves as
undulations on the string induced background, a characteristic feature of these
oscillations being the small ratio of the peak to trough height 

It is interesting that we can find models which illustrate both extremes:
oscillations modulated by a small background and undulations on a
non-oscillatory background. Since we have argued in this paper that none of
these possibilities can be ruled out just by purely theoretical arguments, this
begs the question how can we rule out or constrain these mixed scenarios. The
obvious answer would be to include the parameter $\alpha$ into any analysis of
the CMB data. If we assume, on the basis of the observations of CDM
fluctuations on large scales, that at least some of the fluctuations are
adiabatic, then it would be interesting to see how close to one $\alpha$
is. Therefore, it would be an interesting exercise to investigate the potential
sensitive of the forthcoming satellite missions to this, and since we have
already commented that under some circumstances the inclusion of a defect
background can mimic the effects of other cosmological parameters, it appears
to be likely that this will introduce yet further degeneracies in the
cosmological parameters, which would need to be broken by other measurements. 

Although the power spectra that we have discussed extensively in this paper are
likely to be most well studied aspects of the future datasets, probably the
most characteristic signature of any kind of topological defect is
non-Gaussianity. Clearly, the inclusion of an adiabatic component to the power
spectrum will make the detection of any kind of non-Gaussianity more difficult,
since it will lead to a Gaussian background which would need to be removed
before the test for non-Gaussianity is performed. This will no doubt require
more sophisticated and robust tests than are available at the moment, but if
these mixed perturbation models are seen to be consistent with future data,
then the most compelling argument for their validity over a pure adiabatic
models, will be the detection of these non-Gaussian signatures. 

\section{Discussion and conclusions}
\label{sec-disccon}
 
In this paper we have justified the consideration of models which have two
components to the primordial fluctuations, one created by quantum fluctuations
during inflation and the other due to a network of evolving topological
defects. In places the treatment is totally general, but we have in mind the
idea that the production of defects takes place at the end of inflation. We
have considered general inflationary models with a constant spectral index and
also a specific model based on Supergravity which has some very interesting
properties, including a spectral index which varies with scale. 

In the general case, the inclusion of the string component has some
consequences which are similar to those of a tensor component in the CMB
anisotropies. In that case the inclusion of an extra component in the CMB allows
the matter power spectrum to be pushed down and hence the values of $\sigma_8$
reduced. However, such models also have problems with large scale power: the
SCDM scenario has just about the right amount of power on large scales, but if
the spectrum is pushed down substantially then this feature will be lost. These
mixed perturbation models also have the same problem and it appears that it is
not possible to simultaneously fix the value of $\sigma_8$ and the shape of the
power spectrum on large scales in a universe in which the CDM has a density
close to critical matter, even with extreme values of $n$. It is, nonetheless,
interesting that such models can be made compatible with the simple test of
comparing the amplitude of temperature anisotropies measured by COBE and
measurements of $\sigma_8$. 

If one allows for the inclusion of HDM or a non-zero cosmological constant,
pure adiabatic models can be fit the observed CDM power spectrum without the
need for the inclusion of a component created by strings. However, such
scenarios require anti-biasing relative to the IRAS galaxies which are often
assumed to be good tracers of the CDM. We have suggested that this may achieved
by the inclusion of a string induced component, although it could equally well
be achieved by the inclusion of a tensor component to anisotropy.  

The analogy with the inclusion of tensor component to the anisotropy is not exact, since the strings also contribute to the matter power spectra, albeit at a very low level on large scales. However, on small scales this can have some interesting effects in particular kinks in the power spectrum and substantial increases in power on very small scales. There is some preliminary evidence that such features may exist and the future large scale redshift surveys, such as SDSS and 2Df, will hopefully able to pin this down more accurately. One interesting consequence of the very different behaviour of the spectrum on small scales might be that the number high redshift objects, such as damped Lyman-$\alpha$ systems, might be increased  in scenarios such as CHDM which underproduce such systems in the pure adiabatic limit.

In the case of the Linde-Riotto model we saw that if just adiabatic fluctuations were included then the power spectra were wieldy at odds with the observations. However, when the string induced component is included the power spectra are much more acceptable, although the inclusion of either HDM or non-zero cosmological constant is  necessary to make the shape of the spectrum exactly fit the data.

In whatever inflationary model one considers, these mixed perturbation scenarios have distinct signatures in the CMB. If the adiabatic perturbations dominate, as would have to be the case for a single constant spectral index, the inclusion of a defect component would lead to the modulation of the peak structure, mimicking the effects of baryon drag. While in the models such as that proposed by Linde and Riotto, one finds that the defect component can dominate the CMB anisotropies. In this case the adiabatic oscillations manifest themselves as undulations on the defect spectrum which otherwise has no oscillations. Finally, we suggested that it is a challenge to the forthcoming satellite experiments to confirm or constrain these kinds of scenarios by taking $\alpha$ as a free parameter to need computed. Since there is no a priori reason to believe that these mixed perturbation scenarios can be excluded on any theoretical grounds, such a test would provide a useful information on the nature of physics at high energies.  

\section*{Acknowledgments}

We thank U. Seljak and M. Zaldarriaga for the use of CMBFAST, and
A. Albrecht, J. Magueijo, A. Liddle, A. Linde, N. Turok and R. Jeanerrot for
helpful conversations. The model for strings used in this work was developed by
RAB in collaboration with A. Albrecht and J. Robinson. We would like to thank
them for permission to use these computations in this work. The computations
were done at the UK National Cosmology Supercomputing Center, supported by
PPARC, HEFCE and Silicon Graphics/Cray Research. The work of RAB was funded by
Trinity College and that of JW is supported by a DAAD fellowship HSP III
financed by the German Federal Ministry for Research and Technology. 

During the final stages of this work, we became aware of two other papers which
discussed similar ideas~\cite{CHMb,ACMb}. Ref.~\cite{CHMb} follows up the
suggestion~\cite{jean} that a substantial string induced component would be created in D-term inflation scenario, while ref.~\cite{ACMb} discusses the possibility of string formation as a consequence of open inflation
scenarios~\cite{vilopen}. Both these works reiterate our general conclusions,
but emphasize different aspects. 

\def\jnl#1#2#3#4#5#6{\hang{#1, {\it #4\/} {\bf #5}, #6 (#2).} }
\def\jnltwo#1#2#3#4#5#6#7#8{\hang{#1, {\it #4\/} {\bf #5}, #6; {\it
ibid} {\bf #7} #8 (#2).} } 
\def\prep#1#2#3#4{\hang{#1, #4.} } 
\def\proc#1#2#3#4#5#6{{#1 [#2], in {\it #4\/}, #5, eds.\ (#6).}}
\def\book#1#2#3#4{\hang{#1, {\it #3\/} (#4, #2).} }
\def\jnlerr#1#2#3#4#5#6#7#8{\hang{#1, {\it #4\/} {\bf #5}, #6 (#2).
{Erratum:} {\it ibid.} {\bf #7}, #8.} }
\def\prl{Phys.\ Rev.\ Lett.}
\def\pr{Phys.\ Rev.}
\def\pl{Phys.\ Lett.}
\def\np{Nucl.\ Phys.}
\def\prp{Phys.\ Rep.}
\def\rmp{Rev.\ Mod.\ Phys.}
\def\cmp{Comm.\ Math.\ Phys.}
\def\mpl{Mod.\ Phys.\ Lett.}
\def\apj{Ap.\ J.}
\def\apjl{Ap.\ J.\ Lett.}
\def\aap{Astron.\ Ap.}
\def\cqg{Class.\ Quant.\ Grav.} 
\def\grg{Gen.\ Rel.\ Grav.}
\def\mn{MNRAS}
\def\ptp{Prog.\ Theor.\ Phys.}
\def\jetp{Sov.\ Phys.\ JETP}
\def\jetpl{JETP Lett.}
\def\jmp{J.\ Math.\ Phys.}
\def\zpc{Z.\ Phys.\ C}
\def\cupress{Cambridge University Press}
\def\pup{Princeton University Press}
\def\wss{World Scientific, Singapore}
\def\oup{Oxford University Press}

\pagebreak
\pagestyle{empty}

\end{document}